 \documentclass[preprint2]{aastex}

\shorttitle{}
\shortauthors{Richer et al.}

\begin{document}


\title{The Discrepant Kinematics of ORLs and CELs in NGC 7009 as a Function of Ionization Structure
\footnote{Based on observations made with the European Southern Observatory telescopes obtained from the ESO/ST-ECF Science Archive Facility.}}


\author{Michael G. Richer$^1$, Leonid Georgiev$^{2\,}$\altaffilmark{\dag}, Anabel Arrieta$^3$, and Silvia Torres-Peimbert$^2$}

\affil{$^1$ Instituto de Astronom\'\i a, Universidad Nacional Aut\'onoma de M\'exico, Apartado Postal 106, 22800 Ensenada, BC, M\'exico; richer@astrosen.unam.mx\\
$^2$ Instituto de Astronom\'\i a, Universidad Nacional Aut\'onoma de M\'exico, Apartado Postal 70 - 264, M\'exico, DF, CP 04510; silvia@astroscu.unam.mx\\
$^3$ Universidad Iberoamericana, Departamento de F\'\i sica y Matem\'aticas, Prolongaci\'on Paseo de la Reforma 880, Lomas de Santa Fe, CP 01210 M\'exico DF, M\'exico; anabel.arrieta@ibero.mx}
\altaffiltext{\dag}{deceased, 2012 Dec 26}



\begin{abstract}

We present spatially- and velocity-resolved echelle spectroscopy for NGC 7009 obtained with the UVES spectrograph at the European Southern Observatory's Very Large Telescope.  Our objective is to analyze the kinematics of emission lines excited by recombination and collisions with electrons to determine whether similarities or differences could be useful in elucidating the well-known abundance discrepancy derived from them.  We construct position-velocity maps for recombination, fluorescence, charge transfer, and collisionally-excited lines.  We find a plasma component emitting in the \ion{C}{2}, \ion{N}{2}, \ion{O}{2}, and \ion{Ne}{2} recombination lines whose kinematics are discrepant:  They are incompatible with the ionization structure derived from all other evidence and the kinematics derived from all of these lines are unexpectedly very similar.  We find direct evidence for a recombination contribution to [\ion{N}{2}]$\lambda$5755.  Once taken into account, the electron temperatures from [\ion{N}{2}], [\ion{O}{3}], and [\ion{Ne}{3}] agree at a given position and velocity.  
The electron densities derived from [\ion{O}{2}] and [\ion{Ar}{4}] are consistent with direct imaging and the distribution of hydrogen emission.  
The kinematics of the \ion{C}{2}, \ion{N}{2}, \ion{O}{2}, and \ion{Ne}{2} lines \emph{does not} coincide with the kinematics of the [\ion{O}{3}] and [\ion{Ne}{3}] forbidden emission, indicating that there is an additional plasma component to the recombination emission that arises from a different volume from that giving rise to the forbidden emission from the parent ions within NGC 7009.  Thus, the chemical abundances derived from either type of line are correct only for the plasma component from which they arise.  Apart from [\ion{N}{2}]$\lambda$5755, we find no anomaly with the forbidden lines usually used to determine chemical abundances in ionized nebulae, so the abundances derived from them should be reliable for the medium from which they arise.

\end{abstract}



\keywords{cosmological parameters --- Galaxies: abundances --- ISM: abundances --- ISM: kinematics and dynamics --- planetary nebulae: NGC 7009}


\section{Introduction}\label{section_introduction}

Ionized gas is commonly used to determine chemical compositions and study the enrichment of heavy elements throughout the universe.  To convert the observed line intensities into ionic abundances, the electron temperature and density are normally required \citep[e.g.,][]{aller1984, osterbrock1989}.  Traditionally, the line intensities used to measure the electron density and temperature as well as determine the chemical composition are collisionally-excited forbidden lines of common heavy elements (carbon, nitrogen, oxygen, neon, sulfur, and argon).  Though challenging because they are weak, recombination lines may also be used to determine chemical abundances, as was first attempted by \citet{wyse1942}.  Invariably, the abundances derived from recombination lines are larger than those derived from collisionally-excited lines, at least for carbon, nitrogen, oxygen, and neon \citep{liu2010}, leading to what is known as the abundance discrepancy.  \citet{peimbertpeimbert2006}, \citet{liu2006, liu2010}, and \citet{bohigas2009} provide recent reviews.

The ratio of abundances derived from optical recombination lines (ORLs) and collisionally-excited lines (CELs) is known as the abundance discrepancy factor (ADF).  The ADF is found in both planetary nebulae and \ion{H}{2} regions.  \citet{garciarojasesteban2007} emphasize that the nature of the ADF is, possibly, different in \ion{H}{2} regions, where it is almost always near a factor of two, and in planetary nebulae, where much larger values are encountered about 20\% of the time.  In both \ion{H}{2} regions and planetary nebulae, the ADFs for carbon, nitrogen, oxygen, and neon tend to be similar for a given object, i.e., element ratios such as $\mathrm N/\mathrm O$ are unaffected \citep[e.g.,][]{liu2010}.  

Attempts to explain the ADF phenomenon have focused mainly upon the appropriate temperature to use when deriving abundances and the effects of deviations about some mean temperature.  \citet{peimbert1967} first noted that the use of a single temperature derived from collisionally-excited forbidden lines was likely to overestimate the true mean temperature and underestimate the ionic abundances subsequently derived.  Taking temperature fluctuations into account allows, in many cases, to reconcile the abundances derived from CELs and ORLs \citep[e.g., ][]{estebanetal2004}.  \citet{liuetal2000}, inspired by the unusually large ADF found for NGC 6153, argued that a second plasma component, consisting of cold, hydrogen-deficient clumps, embedded in the normal nebular material was a more natural explanation.  Other ideas have been less explored, but include enhanced radiative or dielectronic recombination and departures of the electron energy distribution from the Maxwellian distribution \citep[e.g.,][]{garnettdinerstein2001, rodriguezgarciarojas2010, pradhanetal2011, nichollsetal2012}.  Dust discs around the central stars are another possibility \citep{bilikovaetal2012}.  Finally, detailed observations of the Orion nebula indicate that density structures, such as proplyds or shocks, can distort the electron temperatures and densities derived from CELs if not taken into account \citep{mesadelgadoetal2008, mesadelgadoetal2009a, mesadelgadoetal2009b, mesadelgadoetal2012, tsamisetal2011}.

There are several means to determine the electron temperature in ionized plasmas \citep[for details, see][]{aller1984, osterbrock1989}.  The result's sensitivity to the underlying temperature and any variations will depend upon the ion used, the conditions in which it is found, and the means by which the transitions are excited.  The most commonly-used temperatures are determined from CELs, the Balmer jump, and the ratio of ORLs.  The intensity of CELs have an exponential dependence upon the electron temperature that characterizes the Maxwellian distribution of the kinetic energy of the free electrons.  Since temperature measurements depend upon the ratio of two such lines, the result will also have an exponential dependence upon temperature.  On the other hand, the intensity of ORLs usually have a power law dependence upon the temperature, $T^{-\beta}$ with typically $\beta\sim 1$.  Ratios of ORLs will therefore have only a weak power law dependence upon the temperature.  The Balmer jump depends upon temperature as $T^{-3/2}$, so the temperature, based upon the ratio of a Balmer line to the Balmer jump, has an approximately square root dependence upon temperature.  

As might be expected as a result of the temperature structure in any given object, the different temperatures do not always agree.  In general, the temperatures derived from CELs are the highest, followed by Balmer jump temperatures, and with temperatures derived from ORLs of neutral helium or ionized oxygen being the coldest \citep[e.g.,][]{liu2010}.  Using hotter temperatures to derive chemical abundances from ORLs helps reconcile them with the abundances determined from CELs.  Obviously, meaningful ionic abundances are obtained only if the electron temperature is appropriate for the emitting ions.  

Much effort has been invested in searching for trends between the ADF and the nebular properties in the hope of uncovering the ADF's origin.  In planetary nebulae, the ADF increases as the difference between the Balmer jump and [\ion{O}{3}]$\lambda\lambda$4363,4959,5007 temperatures increases \citep{liuetal2001}.  Likewise, the ADF increases in planetary nebulae that are larger, more diffuse, and of lower surface brightness, but not with expansion velocity, electron temperature, $\mathrm{He}^{2+}/\mathrm{He}^+$ ratio, or the stellar temperature or luminosity \citep{robertsontessigarnett2005}.  The trends with size, density, and surface brightness could suggest that part of the ADF phenomenon in planetary nebulae has to do with the evolutionary stage.  \citet{tsamisetal2008} find trends with opposite signs between the electron temperature (CELs) and abundances of oxygen derived from ORLs and CELs, finding high recombination line abundances in regions that are more highly ionized and of higher electron temperature.  In \ion{H}{2} regions, there may be weak correlations between increasing ADF and an increasing difference between the Balmer jump and [\ion{O}{3}]$\lambda\lambda$4363,4959,5007 temperatures and with increasing excitation energy, but, within observational uncertainties, no correlation has been found with the abundance of oxygen or of doubly ionized oxygen, the ionization degree, nebular temperatures, the width of the hydrogen lines, or the effective temperature of the main ionizing stars \citep{garciarojasesteban2007}.  

Here, we study NGC 7009, whose ADF is about 5 \citep{liuetal1995,luoetal2001,fangliu2012}.  \citet{rubinetal2002} attempted to observe the temperature fluctuations postulated by \citet{peimbert1967} using imaging and spectroscopy from the Hubble Space Telescope.  This is a difficult endeavor, as any observation will necessarily average the temperature structure along the line of sight to some extent, even at high spectral resolution because of Doppler broadening and the consequent mixing from plasma at different positions (or velocities) cannot be avoided.  \citet{rubinetal2002}  were unable to observe fluctuations as large as needed, but could not exclude the possibility that larger fluctuations existed and were averaged out.  

Here, we present spatially- and velocity-resolved echelle spectroscopy for NGC 7009.  In \S \ref{sec_observations}, we describe the data and the construction of the position-velocity maps.  In \S \ref{sec_results}, we use these maps to investigate the ionization structure of NGC 7009, the contribution of recombination to the [\ion{N}{2}]$\lambda$5755 emission, the component structure of \ion{C}{2} $\lambda$4267, and fluorescence in \ion{N}{2} lines.  
In \S \ref{section_discussion}, we consider the location and possible origins of the ORLs and its consequences for the abundance discrepancy.    
We present our conclusions in \S \ref{section_conclusions}.

\section{Observations and Construction of PV Diagrams}\label{sec_observations}

\begin{figure}  
\includegraphics[width=0.45\columnwidth]{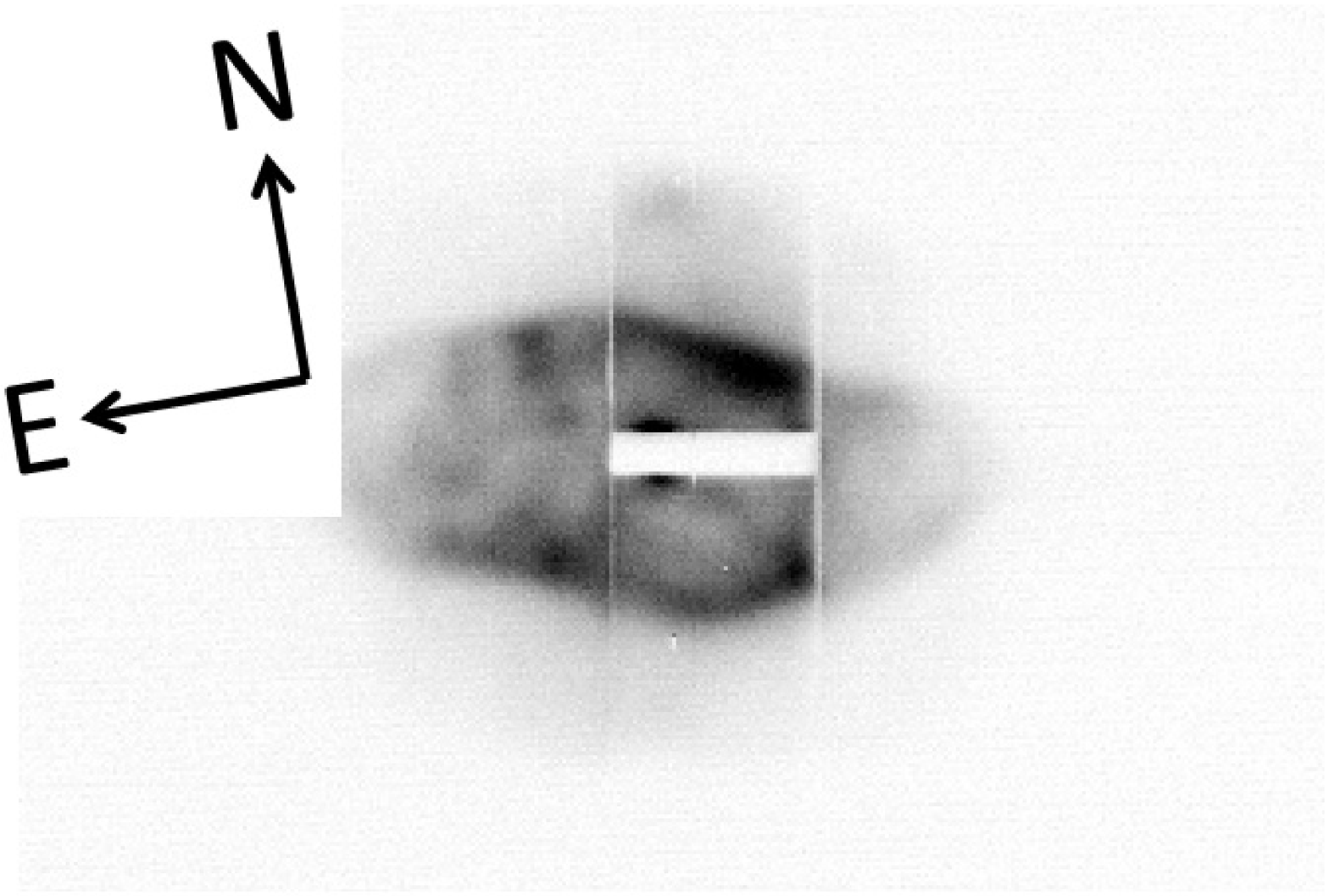}
\includegraphics[width=0.45\columnwidth]{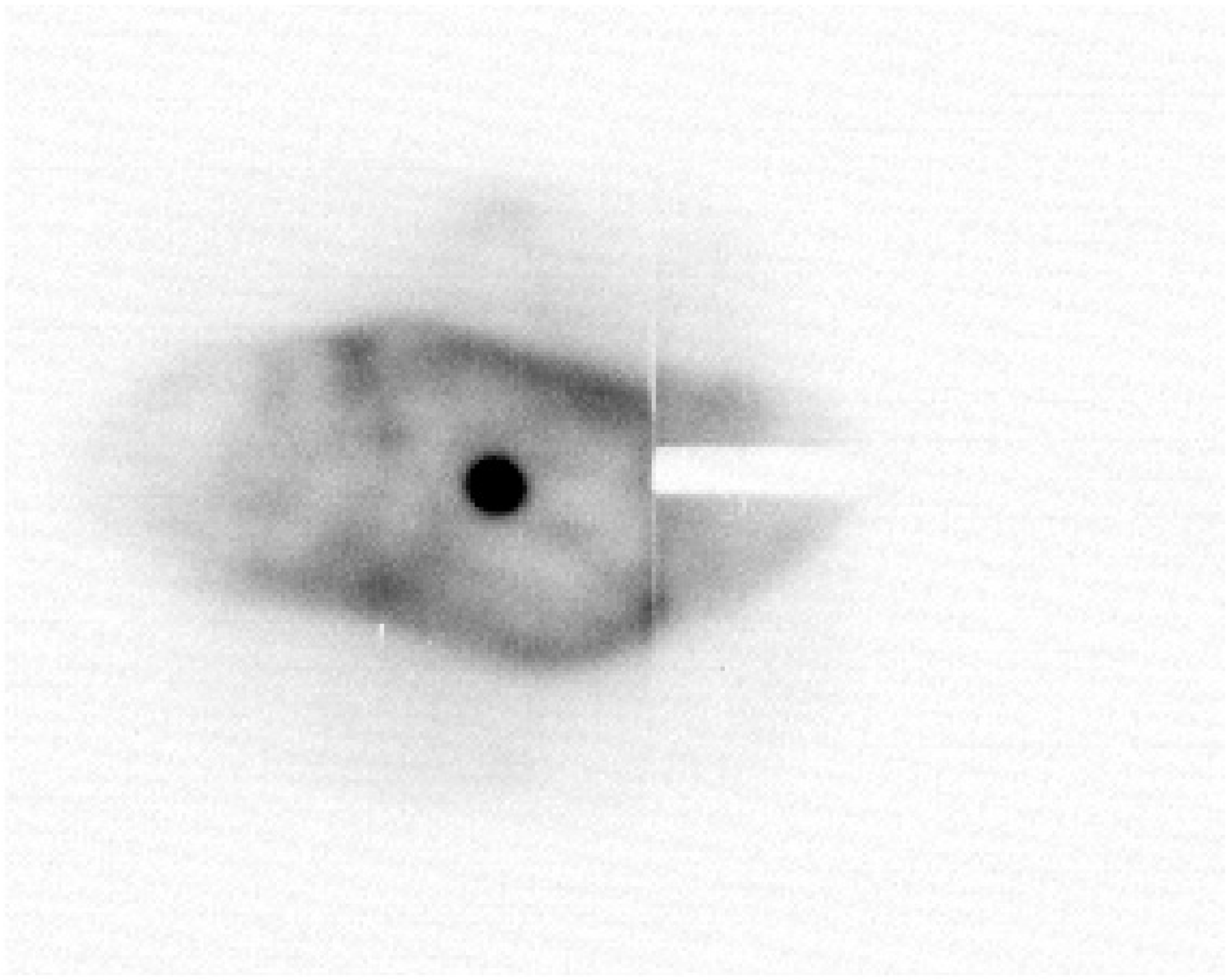}\\
\includegraphics[width=0.45\columnwidth]{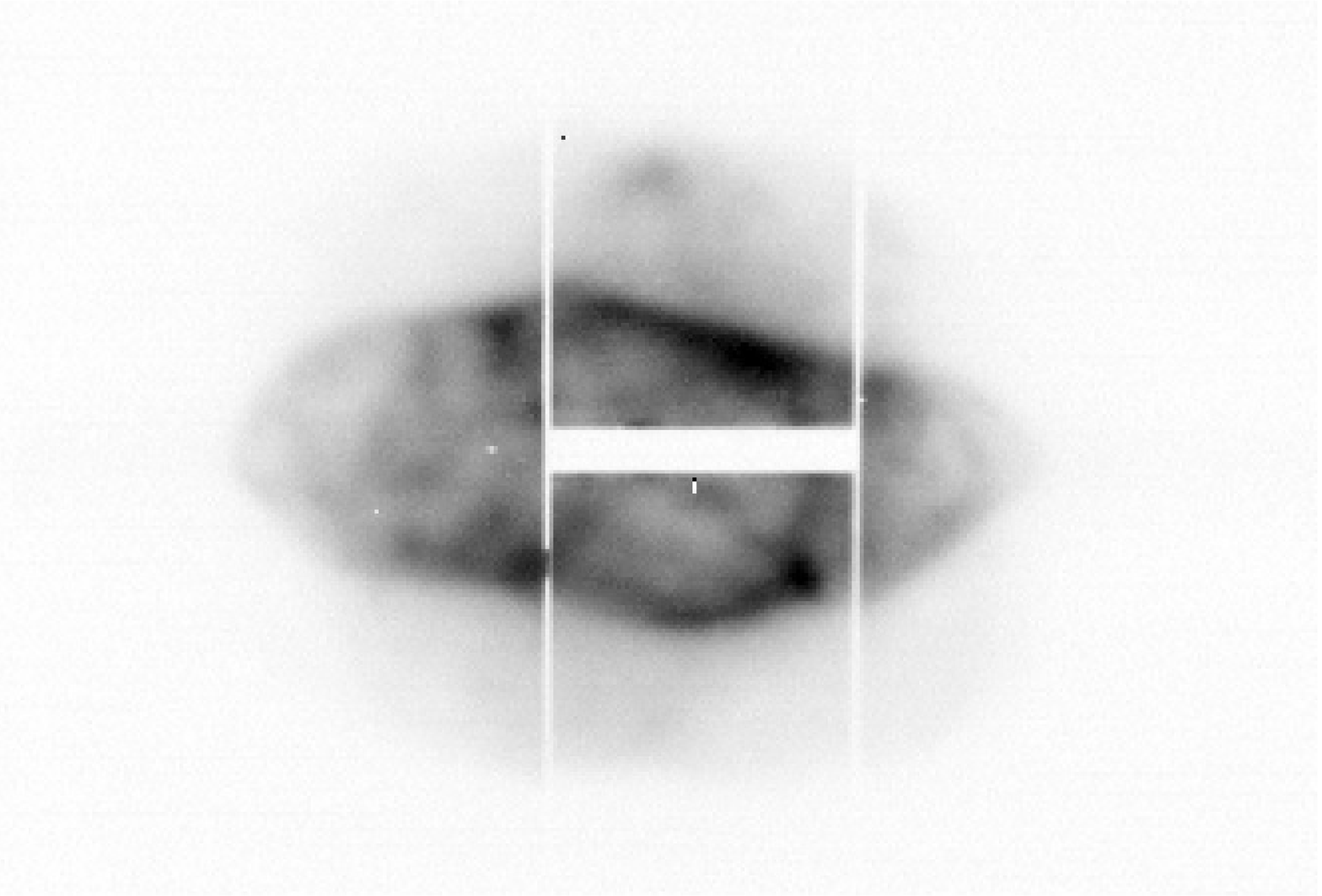}
\includegraphics[width=0.45\columnwidth]{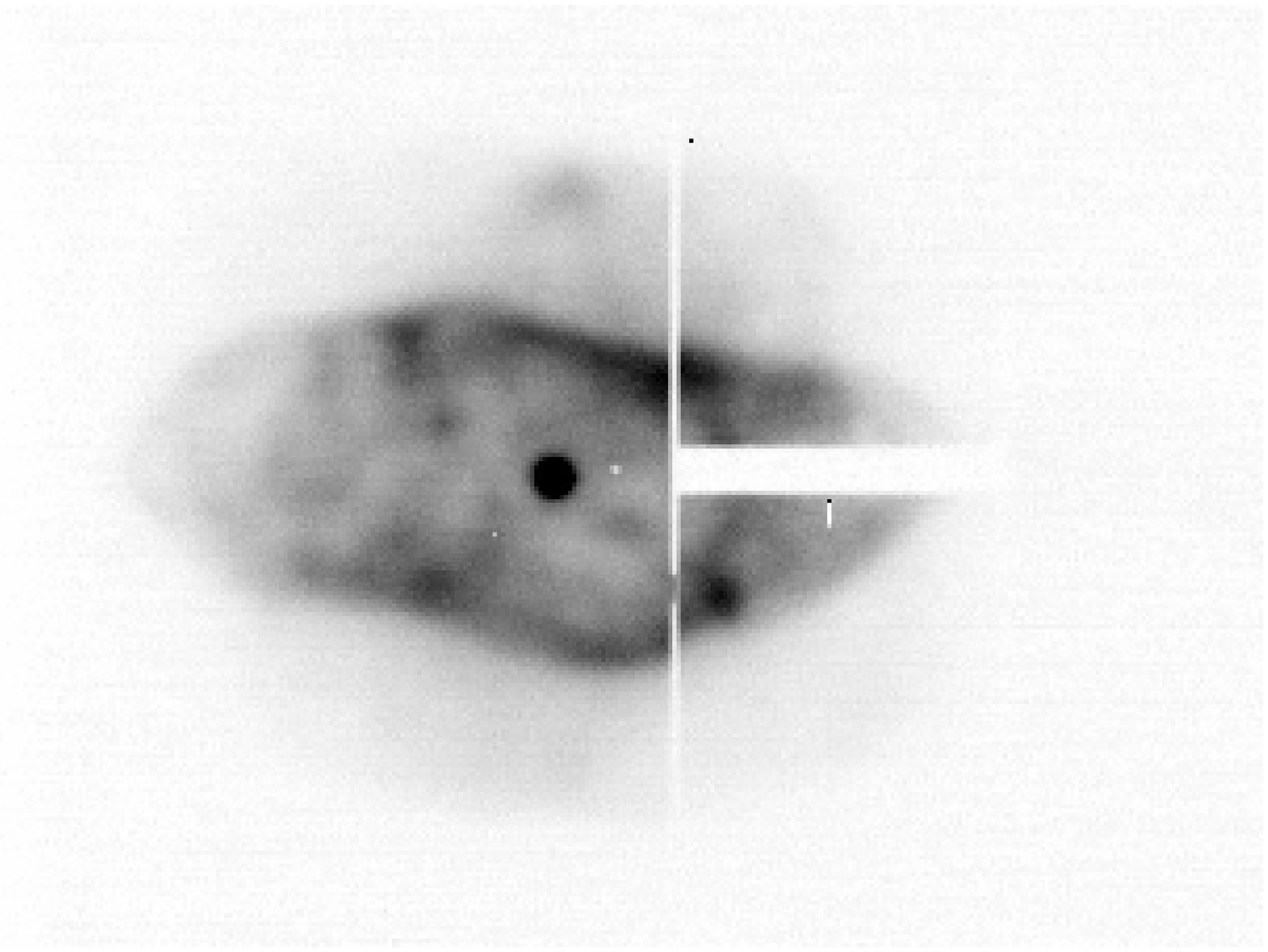}\\
\caption{
These images show the slit positions on NGC 7009, with the blue and red slit positions shown in the top and bottom rows, respectively.  The first slit position is on the left and the second on the right.  The blue slit covers $8\arcsec \times 1\farcs5$ while the red slit covers an area of $11\arcsec \times 1\farcs5$.  The orientation for all panels is shown at upper left.
}
\label{fig_slit_positions}
\end{figure}

The data were retrieved from the ESO data archive from spectra obtained on 2002 Aug 4 using the UV-Visual Echelle Spectrograph (UVES) attached to the European Southern Observatory (ESO) Very Large Telescope (VLT) UT2/Kueyen 8.2m telescope as part of program ID 69.D-0413(A).  Fig. \ref{fig_slit_positions} shows the two slit positions in NGC 7009 analyzed here.  All of the data were acquired within a time span of one hour.  The standard star observed was HR 9087 (program ID 60.A-9022(A)).  

UVES is a two-armed, cross-dispersed echelle spectrograph installed on the Kueyen Nasmyth B platform.  The blue arm uses a single EEV 44-82 $2048\times 4096$ CCD with 15\,$\mu$m pixels while the red arm uses a mosaic of an EEV CCD (same model) and an MIT-LL CCID-20 $2048\times 4096$ CCD with 15\,$\mu$m pixels.  The data were obtained in the standard DIC1 mode with cross dispersers CD2 and CD3 in the blue and red arms, respectively, so the spectra cover the 3259-4518\AA, 4584-5600\AA, and 5655-6672\AA\ wavelength regions.  A 1\farcs5 wide slit was used for both the red and blue arms.  The slit lengths were different, though, 8\farcs0 for the blue arm and 11\farcs0 for the red (Fig. \ref{fig_slit_positions}).  The resulting spectral resolution was approximately 33,000 in the blue, gauged by the width of arc lamp lines, and approximately 31,000 in the red, as gauged from the width of the telluric [\ion{O}{1}]$\lambda$5577 line.  The plate scale was 0\farcs25/pixel and 0\farcs18/pixel in the blue and red arms, respectively.  UVES's atmospheric dispersion corrector (ADC) was used in AUTO mode for all observations of NGC 7009, but was not used for the observation of HR 9087.  

The data were reduced using the Image Reduction and Analysis Facility\footnote{IRAF is distributed by the National Optical Astronomical Observatories, which is operated by the Associated Universities for Research in Astronomy, Inc., under contract to the National Science Foundation.} (IRAF).  We subtracted biases from all images.  We traced the position of the standard star spectrum to determine the positions of the spectral orders.  We normalized each order of the flat field image and inspected the resulting image for interference fringes or spatial variations at the wavelengths of the emission lines of interest.  Since neither were found at the level of the noise, we decided not to apply the flat field correction to the data.  We calibrated the wavelength scale using arc spectra taken the same night, whose internal precision is better than 1\,km/s.  We used the observation of the standard star to calculate the sensitivity function (including the correction for the blaze function).  No correction was applied for reddening, since absolute line ratios will not be crucial to our analysis.  We applied no heliocentric correction to the radial velocity.

\begin{figure*}  
\begin{center}
\includegraphics[width=2.15\columnwidth]{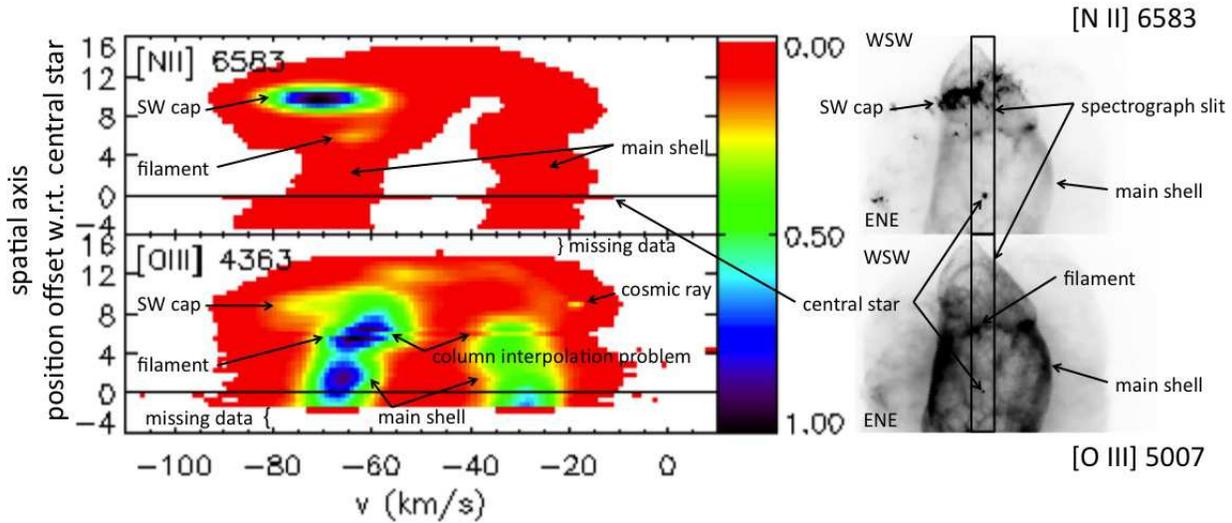}   
\end{center}
\caption{
We present PV diagrams and images for NGC 7009 with the same spatial scale to illustrate their correspondence.  The HST WFPC2 images in [\ion{N}{2}] $\lambda$6583 (top) and [\ion{O}{3}] $\lambda$5007 (bottom) indicate the combined position of the UVES slits (spatial directions indicated).  For the PV diagrams, the spatial axis is the vertical direction and the velocity axis has the most negative velocities (blue-shifted) on the left.  The color scale spans the range of 0\%-100\% of the maximum intensity in both PV diagrams.  The white level extends to 1\% of the maximum intensity.  The horizontal black line in the PV diagrams indicates the position of the central star.  Morphological structures from \citet{sabbadinetal2004} are indicated.  In the [\ion{N}{2}] $\lambda$6583 PV diagram (left), the brightest feature is the SW cap, the filament is considerably fainter, and the main shell is very faint.  There is faint remnant emission from the central star.  Except for the central star, the same emission components are visible in the [\ion{O}{3}] $\lambda$4363 PV diagram (right), but the relative intensities differ, in agreement with their intensities in the direct images, e.g., the filament is now brightest.  We choose this PV map rather than [\ion{O}{3}] $\lambda\lambda$4959,5007 to show the smaller spatial coverage of the UVES blue arm as well as the interpolation problem between the two slit positions that affects all of the blue spectra (and apparently splits the filament).  From the PV diagrams, we infer that the SW cap and filament are on the side of the nebula closest to Earth (blue-shifted).  
}
\label{fig_pv_explanation}
\end{figure*}

We construct position-velocity (PV) maps (or diagrams) from our echelle spectra.  In these diagrams (see Fig. \ref{fig_pv_explanation}), one axis corresponds to the spatial coordinate along the slit (the vertical direction in this case) while the other presents the velocity distribution of the emission at each spatial position.  From PV diagrams such as Fig. \ref{fig_pv_explanation}, it is possible to infer which features seen in direct images are on one side or the other of the object and how others, such as the main shell, are coherent, large-scale spatial constructs that constitute the fundamental three-dimensional structure of the object. 

To construct PV maps for each slit position, we used the spatial position of the standard star as our reference spatial position.  We then extracted pixel-by-pixel strips parallel to the traced position of the standard star.  These spatial strips were then calibrated in both wavelength and flux.  We used IDL routines to resample all of the spectra to a common 0\farcs5 spatial scale and to construct the PV arrays for each position observed in NGC 7009.  In the few cases where it was necessary, cosmic rays were removed using IRAF's \emph{imedit} task.  Since the ADC was not used for the observations of HR 9087, we corrected for differential atmospheric refraction by mapping the position of NGC 7009's central star in slit position 1 as a function of wavelength to determine the spatial offset correction.  This correction was applied to the PV arrays for both slit positions in NGC 7009.  We removed the spectra of the central star and the nebular continuum from the PV arrays by interpolating the spatial profiles on both sides of the emission lines of interest.  

Finally, the PV maps for the two slit positions were merged into a single PV map by scaling the flux in the position 2 spectrum to that in the position 1 spectrum in the region of overlap.  Since there is no overlap for the blue spectra, the flux scaling is based upon the red spectra only.  The lack of overlap in the blue spectra produces an \lq\lq interpolation gap" in the PV maps (see Fig. \ref{fig_pv_explanation}).  Note that the [\ion{O}{3}] $\lambda 4959$ line is saturated in the second slit position, which introduces artifacts when this PV diagram is compared to others.  We have identified this region in the following figures when relevant.  

We present PV diagrams for NGC 7009 in different emission lines in Fig. \ref{fig_pv_diagrams}.  The lines and the maximum intensity are identified at upper left in each panel.  In all cases, the velocity range spans the range from -110 to +10 km/s and is not corrected for heliocentric motion.  The horizontal line indicates the position of the central star and the spatial scale is given in seconds of arc.  The instrumental broadening limits the spectral resolution for some kinematic components of some lines, such as the approaching wall of the main shell in the lines of 
[\ion{O}{3}] $\lambda$4959 and [\ion{Ar}{4}] $\lambda$4711.  For the hydrogen and helium lines thermal broadening and the fine structure (\ion{He}{1} singlets excepted) further broadens the lines.  For other lines, presumably the broadening is due to the kinematic structure.  For all the lines shown in Fig. \ref{fig_pv_diagrams}, the distribution of the emission on the receding side (redshifted) of the nebula is broadened beyond that due to instrumental broadening, presumably due to kinematic structure.  

The main features in our PV diagrams are what \citet{sabbadinetal2004} call the main shell and the SW cap (see Figs. \ref{fig_slit_positions} and \ref{fig_pv_explanation}).  
In Fig. \ref{fig_pv_diagrams}, the main shell is prominent in higher ionization lines, such as \ion{He}{1} $\lambda$4922, [\ion{O}{3}] $\lambda$4959, [\ion{Ar}{4}] $\lambda$4740, or \ion{He}{2} $\lambda$4686, but very faint in low ionization lines, such as [\ion{O}{2}] $\lambda$3726, and likely optically-thin due to the absence of [\ion{O}{1}] $\lambda$ 6300 \citep[see also][]{sabbadinetal2004}.  The SW cap is seen best in the [\ion{N}{2}] $\lambda$6583 image (Fig. \ref{fig_pv_explanation}), where it appears as a bright filamentary structure running nearly perpendicular to the main shell, and is best seen in the [\ion{O}{2}] $\lambda$3726 and [\ion{O}{1}] $\lambda$6300 PV diagrams in Fig. \ref{fig_pv_diagrams}.  Its velocity and low level of ionization, indicate that it is outside the main shell \citep{sabbadinetal2004}.  The SW cap has very low mass since it is nearly absent in the H$\beta$ PV diagram (Fig. \ref{fig_pv_diagrams}), but it is the only direction covered by these observations that could be optically-thick to ionizing radiation (strong [\ion{O}{1}] $\lambda$6300).  

Comparing PV diagrams quantitatively requires care, e.g., taking ratios to compute the physical conditions or to model one PV diagram as a sum of several others, since such comparisons require precise alignment in both the spatial and velocity axes.  We avoid comparing the PV diagrams of lines with fine structure.  Only occasionally must we compare lines with different thermal widths (and account for it).  We require a physically-consistent wavelength scale that places co-spatial emission from different ions at the same radial velocity, while mitigating the effects of any errors due to our wavelength calibration or laboratory wavelength measurements.  The assumptions involved should not introduce spurious kinematic structure.  Our adopted wavelengths are given in Table \ref{tab_wavelengths}, taken from \citet{cleggetal1999} for \ion{H}{1} and \ion{He}{2} lines, \citet{bowen1960} for CELs, and from the NIST database \citep{ralchenkoetal2011}\footnote{http://physics.nist.gov/asd} for the remaining lines.  

There are many studies of the structure and kinematics of NGC 7009 \citep[e.g.,][]{bohigasetal1994,goncalvesetal2003,sabbadinetal2004,steffenetal2009,phillipsetal2010}.  
Its kinematic structure appears reasonably simple, with homologous (radial) expansion dominating except perhaps towards the eastern and western extremities of the shell \citep{sabbadinetal2004, steffenetal2009}.  These results suggest that the mean velocity of the middle section of the main shell may be constant for all ions.  This should be true for lines from a given ion or lines from co-spatial ions.  Whether this assumption is reasonable for lines that are not expected to arise co-spatially is unclear, e.g., \ion{He}{2} $\lambda$4686 and [\ion{O}{2}] $\lambda\lambda$3726,3729, but we are typically able to avoid comparing the PV diagrams of these line pairs.  So, we align all PV diagrams in velocity by measuring the velocities of the blue- and red-shifted components in the middle section of the main shell and shifting the mean of these two velocities to a common mean velocity.  We measure these velocities in the the spatial region from $+0\farcs5$ to $+2\farcs5$ SW of the central star (see Fig. \ref{fig_pv_diagrams}), chosen to avoid the central star, but still very central in order to suffer little from projection effects due to the geometry of nebular shell.  We adopt the mean velocity for the sum of the \ion{O}{2} $\lambda\lambda$4639,4649 lines as the common mean velocity.  
Note that measurements of \emph{line splitting} are unaffected by the velocity alignment of the PV diagrams.  

\section{Results}\label{sec_results}

\subsection{Ionization structure}\label{section_ionization_structure}

\begin{figure*}
\begin{center}
\includegraphics[width=2\columnwidth]{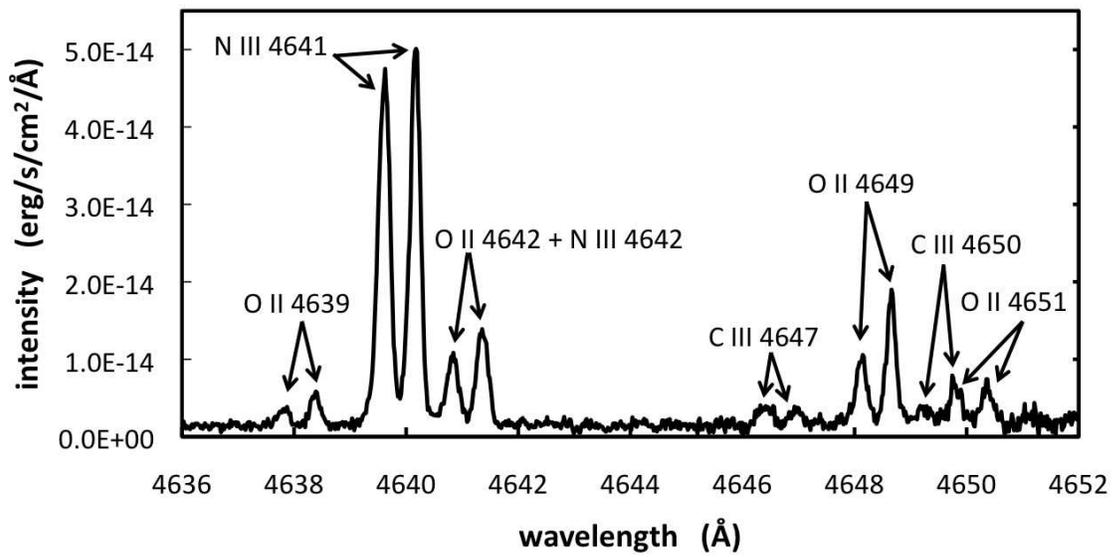}  
\end{center}
\caption{
We present the spectrum around the \ion{O}{2} $\lambda\lambda$4639,4642,4649,4651 lines.  The \ion{O}{2} $\lambda\lambda$4639,4649 lines are not blended with \ion{N}{3} or \ion{C}{3} lines though the \ion{O}{2} $\lambda\lambda$4642,4651 lines are affected.
}
\label{fig_1d_spec_oii}
\end{figure*}

Table \ref{tab_wavelengths} confirms the simple structure of NGC 7009's main shell.  The last column presents the total velocity splitting between the main shell components for the spatial region from $+0\farcs5$ to $+2\farcs5$ SW of the central star.  We determined the total velocity splitting by fitting a single Gaussian to the velocity component for the front and back walls of the main shell.  Most of the lines in Table \ref{tab_wavelengths} are isolated at the resolution of our observations.  Fig. \ref{fig_1d_spec_oii} presents a section of spectrum spanning the \ion{O}{2} $\lambda\lambda$4639,4642,4649,4651 lines that are usually blended with the \ion{N}{3} $\lambda\lambda$4641,4642 Bowen fluorescence lines and the \ion{C}{3} $\lambda\lambda$4647,4650,4651 recombination lines at lower resolution \citep[cf.][]{fangliu2011}.  Clearly, at our resolution, the \ion{O}{2} $\lambda\lambda$4639,4649 recombination lines are free of blending.  Of the other lines in Table \ref{tab_wavelengths}, \ion{N}{3} $\lambda$4379 is perhaps contaminated by \ion{Ne}{2} $\lambda$4379 (13\% level), \ion{He}{1} $\lambda$4922 may be contaminated by [\ion{Fe}{3}] $\lambda$4922 (10\% level), while \ion{C}{2} $\lambda$4267 and [\ion{N}{2}] $\lambda$6583 may be contaminated at the 1\% level \citep{fangliu2011}.  We disregard contamination below 1\%.  

There is clear ionization stratification within the main shell  \citep[e.g., ][Table \ref{tab_wavelengths}, Fig. \ref{fig_pv_diagrams}]{wilson1950}.  The lines expected to be emitted from the bulk of the main shell's volume, the H Balmer lines, \ion{He}{1} $\lambda$4922, and [\ion{O}{3}] $\lambda$4959 have a velocity splitting of 39-40\,km/s.  The lines from the most highly ionized ions, \ion{He}{2} $\lambda$4686, \ion{N}{3} $\lambda$4379, and \ion{O}{3} $\lambda$3707, have the smallest velocity splitting at 33-34\,km/s.  At an intermediate velocity splitting, about 36\,km/s, we find \ion{C}{2} $\lambda$6578, \ion{O}{2} $\lambda\lambda$4639,4649, and [\ion{Ar}{4}] $\lambda\lambda$4711,4740.  The largest velocity splitting, about 41\,km/s, occurs for [\ion{N}{2}] $\lambda$6583 and [\ion{O}{2}] $\lambda\lambda$3726,3729.  
We note that the velocity splitting for [\ion{N}{2}] $\lambda$5755 and [\ion{O}{3}] $\lambda$4363 differs from that for [\ion{N}{2}] $\lambda$6548 and [\ion{O}{3}] $\lambda$4959, respectively.  Thus, within each wall of the main shell, the average velocity of the highest ionization species differs from that of the lowest ionization species by only about 4\,km/s.  The main shell does not emit in [\ion{O}{1}] over the area surveyed here, so we conclude that it is optically thin to hydrogen ionizing radiation over this area.

The lines with intermediate velocity splitting of about 35-37\,km/s divide into two groups.  One group is from ions expected in the region where He$^{2+}$ recombines to He$^+$, i.e., [\ion{Ar}{4}] $\lambda\lambda$4711,4740, the Bowen lines of \ion{O}{3} $\lambda$3444 and \ion{N}{3} $\lambda$4641 excited by \ion{He}{2} Ly $\alpha$ in the innermost part of the O$^{2+}$ zone, and lines excited by the $\mathrm O^{3+} + \mathrm H^{\circ} \rightarrow \mathrm O^{2+} + \mathrm H^+$ charge exchange reaction.  Ionization stratification should locate this emission naturally at velocities between those of \ion{He}{2} $\lambda$4686 and [\ion{O}{3}] $\lambda$4959.  The second group, recombination lines of \ion{C}{2}, \ion{N}{2}, \ion{O}{2}, and \ion{Ne}{2}, is not expected where He$^{2+}$ recombines to He$^+$, but where doubly ionized ions of these elements are most common within the O$^{2+}$/Ne$^{2+}$ zone, and so with a velocity splitting similar to [\ion{O}{3}] $\lambda$4959 and [\ion{Ne}{3}] $\lambda$3869.  

Our data sample the ionization structure of oxygen in NGC 7009 particularly well.  The \ion{O}{3} $\lambda$3707 recombination line samples the O$^{3+}$ zone, the \ion{O}{3} $\lambda\lambda$3444,3757,5592 lines sample the transition zone between O$^{3+}$ and O$^{2+}$ via both Bowen fluorescence and charge transfer, the [\ion{O}{3}] $\lambda\lambda$4363,4959 and \ion{O}{2} $\lambda\lambda$4639,4949 lines sample the O$^{2+}$ zone in collisionally-excited and recombination lines, the [\ion{O}{2}] $\lambda\lambda$3726,3729 lines sample the O$^+$ zone, and the [\ion{O}{1}] $\lambda$6300 line samples the transition zone where both oxygen and hydrogen recombine to neutral atoms.  (The PV diagrams for \ion{O}{3} $\lambda\lambda$3757,5592 are statistically indistinguishable from that for \ion{O}{3} $\lambda$3444, but of considerably lower signal-to-noise.)  
The kinematics are expected to reflect the ionization structure, with the innermost zones showing the smallest line splitting and the outermost zones suffering the greatest splitting.  This expectation is borne out (Table \ref{tab_wavelengths}), 
but for the anomaly already noted, that the velocity-splitting of the \ion{O}{2} recombination lines is smaller than that for the [\ion{O}{3}] $\lambda$4959 line.  

Other ions confirm the ionization structure inferred from oxygen:  \ion{N}{3} $\lambda$4379 coincides with \ion{O}{3} $\lambda$3707, [\ion{Ne}{3}] $\lambda$3869 coincides with [\ion{O}{3}] $\lambda$4959, and [\ion{N}{2}] $\lambda$6583 coincides with [\ion{O}{2}] $\lambda\lambda$3726,3729.  Comparing the PV diagrams for [\ion{O}{3}] $\lambda$4959 and [\ion{O}{2}] $\lambda$3726 differentially with respect to \ion{He}{1} $\lambda\lambda$4922,5016,5876 and \ion{He}{2} $\lambda$4686, we find the expected differences, given the differences in ionization potentials:  [\ion{O}{2}] $\lambda$3726 traces material outside the He$^+$ zone, but also includes the outermost material, in both velocity and spatial extent, traced by He$^+$.  [\ion{O}{3}] $\lambda$4959 generally follows the He$^+$ distribution, but also shows emission at the positions and velocities where the \ion{He}{2} $\lambda$4686 emission is enhanced, reflecting the O$^{2+}$ ions in the outermost part of the He$^{2+}$ zone.  Therefore, the ionization structure of oxygen, as traced by \ion{O}{3} $\lambda$3707, Bowen and charge exchange emission, [\ion{O}{3}] $\lambda$4959, [\ion{O}{2}] $\lambda$3726, and [\ion{O}{1}] $\lambda$6300, appears secure.  This structure is also that predicted by hydrodynamic models of planetary nebulae at NGC 7009's stage of evolution \citep[e.g.,][]{villaveretal2002, perinottoetal2004}.

On the other hand, the anomaly of the \ion{O}{2} lines is not isolated.  The recombination lines of \ion{C}{2}, \ion{N}{2}, and \ion{Ne}{2} have a line splitting similar to that of the \ion{O}{2} recombination lines.  Therefore, all of these recombination lines have velocity splittings that are smaller than expected.  

\subsection{Physical conditions and [\ion{N}{2}] $\lambda$ 5755 recombination}\label{section_physical_conditions}\label{section_n2_5755}

Fig. \ref{fig_pv_conditions} presents PV diagrams for the physical conditions of electron temperature and density.  To compute these PV diagrams, it is necessary to take the ratio of two PV diagrams, e.g., [\ion{O}{3}] $\lambda$4363 and [\ion{O}{3}] $\lambda$4959 to compute the [\ion{O}{3}] temperature or [\ion{Ar}{4}] $\lambda$4711 and [\ion{Ar}{4}] $\lambda$4740 to compute the [\ion{Ar}{4}] density.  
In the case of the [\ion{O}{3}] and [\ion{N}{2}] temperatures, we converted the [\ion{O}{3}] $\lambda\lambda$4363,4959 and [\ion{N}{2}] $\lambda\lambda$5755,6584 line ratios to electron temperatures using equations 5.4 and 5.5, respectively, from \citet{osterbrock1989}, assuming an electron density of 4000\,cm$^{-3}$ and the atomic data used therein \citep{nussbaumerrusca1979, nussbaumerstorey1981}.  For the [\ion{O}{2}] and [\ion{Ar}{4}] density maps, we converted the [\ion{O}{2}] $\lambda\lambda$3726,3729 and [\ion{Ar}{4}] $\lambda\lambda$4711,4740 line ratios to electron density using the atomic data incorporated into the STSDAS version 3.8 nebular package, assuming an electron temperature of 10,000\,K \citep{bowen1960,kaufmansugar1986,zeippenetal1987,mclaughlinbell1993,wieseetal1996}.

\begin{figure*}
\begin{center}
\includegraphics[width=2.15\columnwidth]{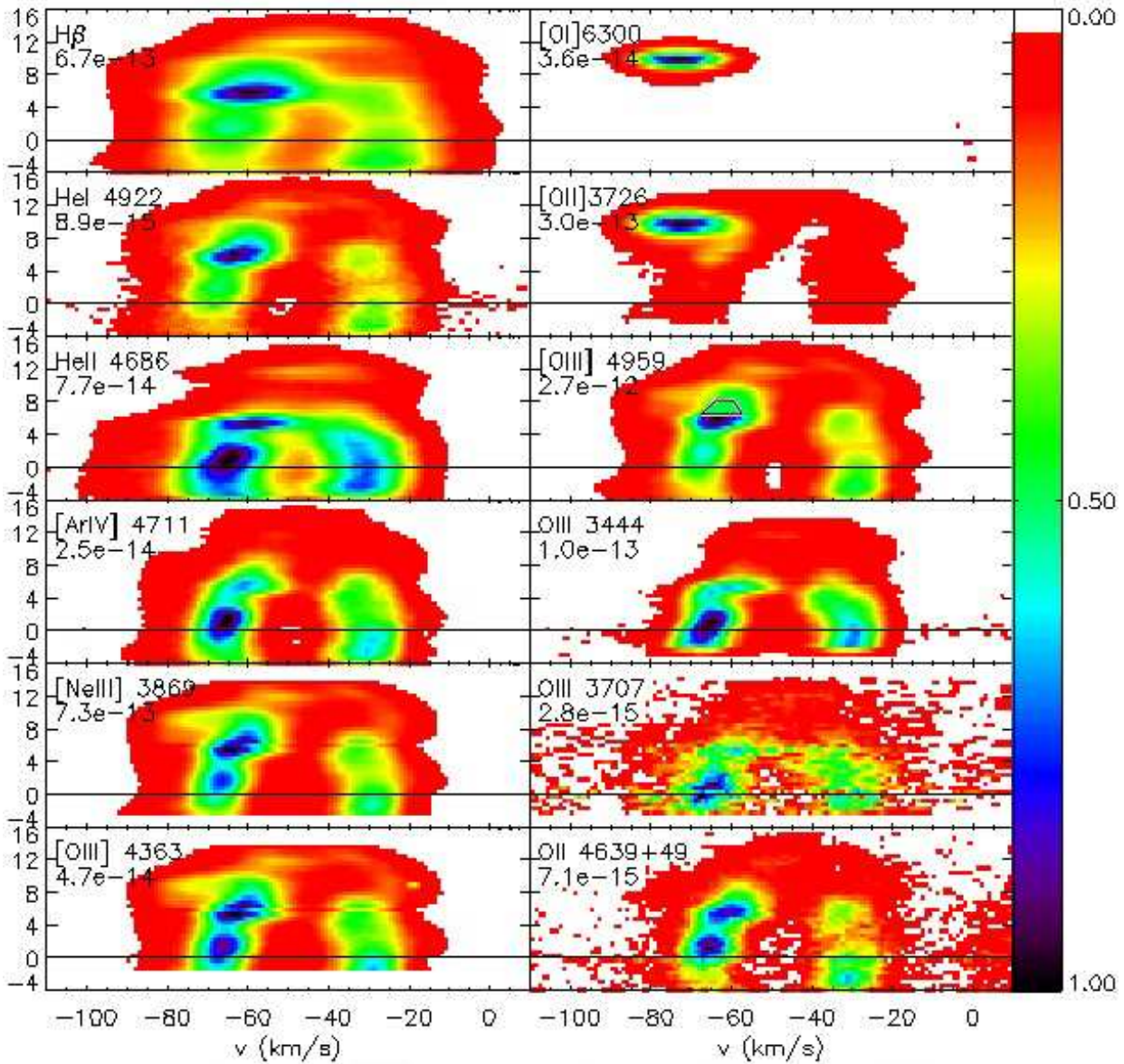}  
\end{center}
\caption{
We present PV diagrams for NGC 7009 in the light of different emission lines.  In each panel, the emission line and maximum intensity are identified in the upper left.  The color scale spans the range of 0\%-100\% of the maximum intensity.  The white level extends to 2\% of the maximum intensity.  The spatial scale is indicated on the left, measured in seconds of arc from the position of the central star (indicated by the horizontal line).  The radial velocity scale spans from -110 to +10 km\,s$^{-1}$ (no heliocentric correction).  The spatial region used to measure the velocity splitting in Table \ref{tab_wavelengths} spans the spatial range from $+0\farcs5$ to $+2\farcs5$, immediately above/SW of the central star.  
The emission at positive velocities in the \ion{O}{2} $\lambda\lambda$4639,4649 PV diagram is due to the nearby \ion{C}{3} $\lambda$4650 line (see Fig. \ref{fig_1d_spec_oii}).  The trapezoid in the [\ion{O}{3}] $\lambda$4959 PV diagram indicates the region (part of the filament; Fig. \ref{fig_pv_explanation}) where the flux in this line is saturated.
}
\label{fig_pv_diagrams}
\end{figure*}

\begin{figure*}
\begin{center}
\includegraphics[width=2.15\columnwidth]{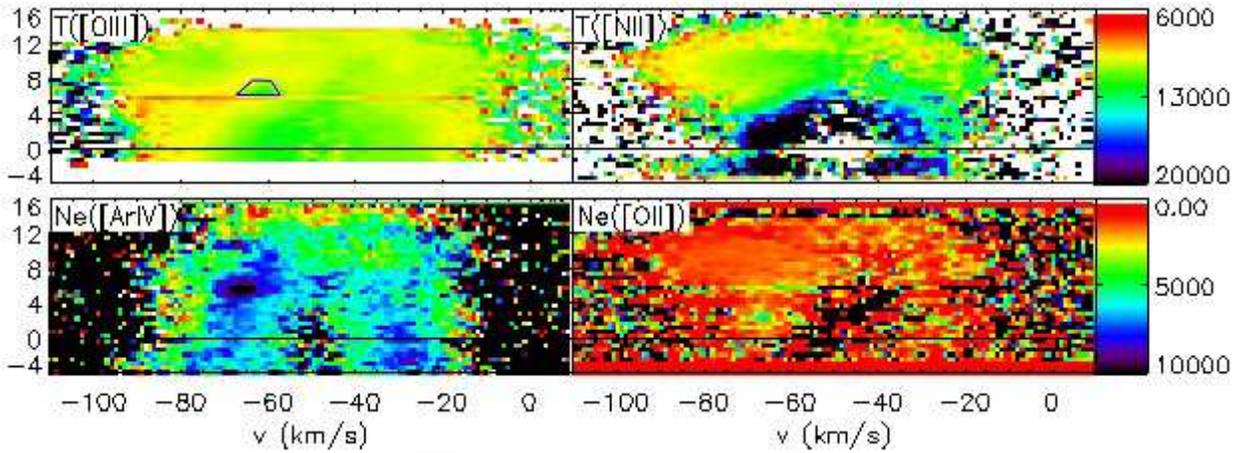}  
\end{center}
\caption{
We present PV diagrams for the [\ion{O}{3}] and [\ion{N}{2}] temperatures (top: left and right, respectively) and the [\ion{Ar}{4}] and [\ion{O}{2}] densities (bottom: left and right, respectively).  The [\ion{O}{3}] and [\ion{N}{2}] temperatures agree in the top part of the PV diagram, with temperatures varying by about 600\,K about a value of 10,000\,K, but differ by about a factor of two in the bottom part of the diagram.  The [\ion{Ar}{4}] density map is quite uniform throughout the main shell, except at the position of the filament (Fig. \ref{fig_pv_explanation}), which has a higher density.  The low ionization material visible in [\ion{O}{2}] $\lambda$3726, including the SW cap (Fig. \ref{fig_pv_explanation}) and that interior to the Ar$^{3+}$ zone is of substantially lower density.  The trapezoid (top left) indicate the saturated region in the [\ion{O}{3}] $\lambda$4959 PV diagram.  The white level extends to 6,390\,K in the top row and to 300 cm$^{-3}$ in the bottom row.  See Fig. \ref{fig_pv_diagrams} for further details.
}
\label{fig_pv_conditions}
\end{figure*}

\begin{figure*}
\begin{center}
\includegraphics[width=2.15\columnwidth]{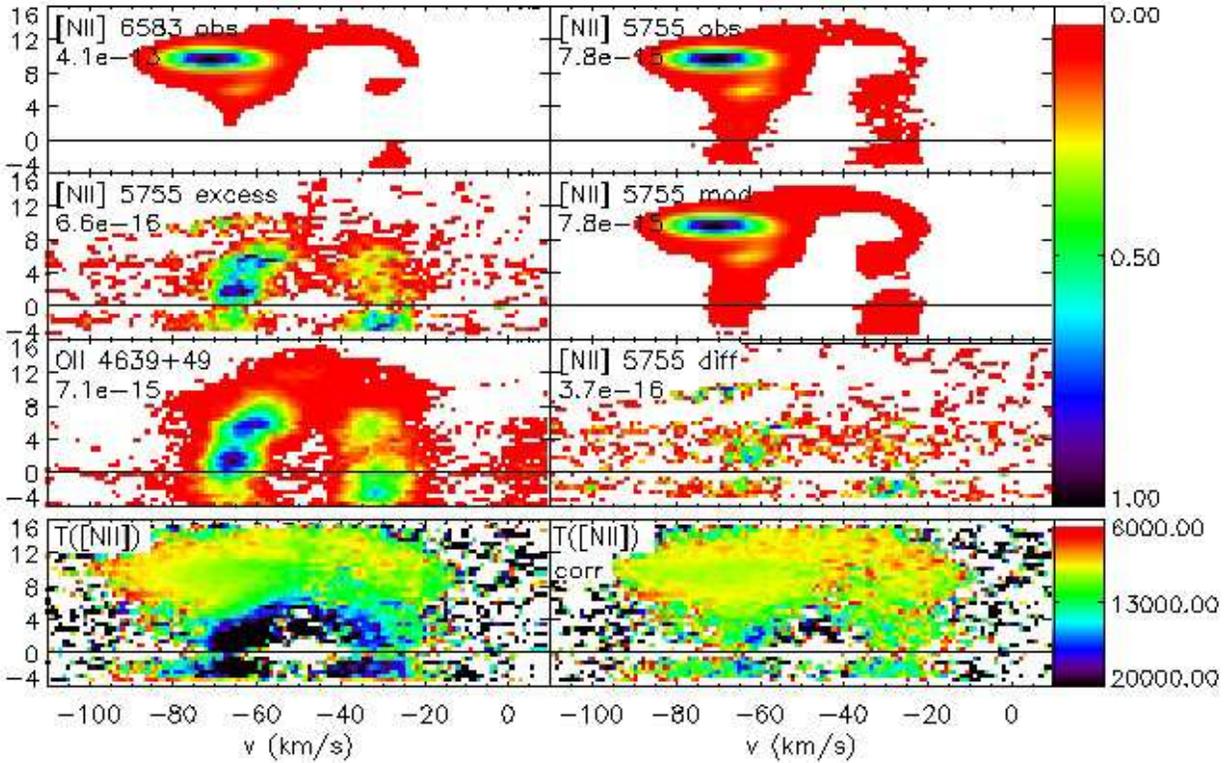}  
\end{center}
\caption{
We present observed and modeled PV diagrams for the [\ion{N}{2}] $\lambda\lambda$5755,6583 and \ion{O}{2} $\lambda\lambda$4639,4649 lines.  The modeled [\ion{N}{2}] $\lambda$5755 map is constructed by adding a scaled \ion{O}{2} $\lambda\lambda$4639,4649 map to the scaled [\ion{N}{2}] $\lambda$6583 map (see text).  The map of the [\ion{N}{2}] $\lambda$5755 excess is obtained by subtracting a scaled [\ion{N}{2}] $\lambda$6583 map from the observed [\ion{N}{2}] $\lambda$5755 map.  The PV diagram of the excess emission is very similar to the \ion{O}{2} $\lambda\lambda$4639,4649 PV diagram.  At right in the third row is the difference of the observed [\ion{N}{2}] $\lambda$5755 map and the modeled map.  The residuals are less than 10\% of the maximum intensity of the [\ion{N}{2}] $\lambda$5755 map, which confirms that the excess [\ion{N}{2}] $\lambda$5755 emission coincides closely with that of \ion{O}{2} $\lambda\lambda$4639,4649.   
In the bottom row are maps of the [\ion{N}{2}] temperature using the [\ion{N}{2}] $\lambda$5755 PV diagram as observed (left) and after subtracting the scaled \ion{O}{2} $\lambda\lambda$4639,4649 PV diagram (right).  The white level extends to 3\% of the maximum intensity in top three rows and to 6,390\,K in the bottom row.  See Fig. \ref{fig_pv_diagrams} for further details.
}
\label{fig_pv_n2}
\end{figure*}

The density distribution agrees with previous findings, e.g., \citet{sabbadinetal2004} with which we can compare most easily.  The electron density in the main shell is rather uniform based upon the [\ion{Ar}{4}] lines, with values typically of $\sim 6000$\,cm$^{-3}$.  The only exception is a dense filament crossing the approaching face of the main shell, seen clearly in the HST images (Fig. \ref{fig_pv_explanation}) and the PV diagrams for H$\beta$, \ion{He}{1} $\lambda$4922, [\ion{O}{3}] $\lambda$4959 and [\ion{Ne}{3}] $\lambda$3869 (Fig. \ref{fig_pv_diagrams}).  Since this filament is fainter in \ion{He}{2} $\lambda$4686, \ion{O}{3} $\lambda$3444, and [\ion{N}{2}] $\lambda$6583 (Figs. \ref{fig_pv_explanation}, \ref{fig_pv_diagrams} and \ref{fig_pv_n2}) than in [\ion{O}{3}] $\lambda$4959, it is presumably located midway through the main shell structure.  The SW cap is less dense, $\sim 2000$\,cm$^{-3}$ as judged from the [\ion{O}{2}] lines.  The lower density makes sense given its near-absence in the H$\beta$ PV diagram (Fig. \ref{fig_pv_diagrams}).   

The electron temperature found in the outer (SW) part of the nebula (upper part of the PV diagrams) fluctuates around a value of $10^4$\,K, with good agreement between the temperatures from the [\ion{O}{3}] and [\ion{N}{2}] lines.  As expected, the temperature is lower in the filament (Fig. \ref{fig_pv_explanation}) where the density is higher.  Both the [\ion{O}{3}] and [\ion{N}{2}] lines indicate that the regions nearer the central star are hotter, but the variation implied by the [\ion{N}{2}] lines is much greater, up to values exceeding $2\times 10^4$\,K.  While we do not have the signal-to-noise to construct a PV diagram for the [\ion{Ne}{3}] temperature, we can calculate its value in the main shell components near the central star (the same region used to measure the kinematics in Table \ref{tab_wavelengths}, 0\farcs5--2\farcs5 SW of the central star).  We find [\ion{Ne}{3}] $\lambda$3869/[\ion{Ne}{3}] $\lambda$3342 ratios of 573 and 556 in the blue- and red-shifted components of the main shell, respectively.  These ratios imply temperatures of 10750\,K and 10840\,K for the blue- and red-shifted components, respectively, and confirm the temperature found from the [\ion{O}{3}] lines in this region, precisely where the [\ion{N}{2}] temperature is anomalous.  

Important differences between the [\ion{N}{2}] and [\ion{O}{3}] temperatures have been found previously in NGC 7009 \citep[e.g.,][]{liudanziger1993,liuetal1995,fangliu2012} and have been attributed to a recombination contribution to [\ion{N}{2}] $\lambda$5755 \citep{liuetal2000}.  In Fig. \ref{fig_pv_n2}, we compare the [\ion{N}{2}] $\lambda\lambda$5755,6583 PV diagrams (top row).  Clearly, there are additional emission components in the [\ion{N}{2}] $\lambda$5755 map.  These could arise via collisional excitation if the regions of excess [\ion{N}{2}] $\lambda$5755 emission were either very hot or very dense.  If these regions were very hot, they should be even more prominent in [\ion{O}{3}] $\lambda$4363 and they are not (Fig. \ref{fig_pv_diagrams}).  If these regions were very dense, this should be evident from the ratio of the [\ion{Ar}{4}] $\lambda\lambda$4711,4740 or [\ion{O}{2}] $\lambda\lambda$3726,3729 lines, if they are ionized, or from the presence of [\ion{O}{1}] $\lambda$6300 if they are neutral, but the [\ion{Ar}{4}] and [\ion{O}{2}] densities are not unusual and [\ion{O}{1}] $\lambda$6300 is absent, so high density can also be ruled out (Figs. \ref{fig_pv_diagrams} and \ref{fig_pv_n2}).  
Therefore, the excess [\ion{N}{2}] $\lambda$5755 emission is presumably not the result of collisional excitation.

If the additional components in [\ion{N}{2}] $\lambda$5755 are due to recombination, it should be possible to model this map as a linear superposition of the [\ion{N}{2}] $\lambda$6583 map and a recombination line map, such as \ion{O}{2} $\lambda\lambda$4639,4649.  (Such a decomposition implicitly assumes a constant electron temperature in the N$^+$ zone.)  In Fig. \ref{fig_pv_n2}, we present a model [\ion{N}{2}] $\lambda$5755 map (second row, right) and it is very similar to the observed [\ion{N}{2}] $\lambda$5755 map, as the difference of the two reveals (third row, right).  (The intercept of the fit was forced through the origin, equivalent to supposing that the continuum emission in all PV diagrams is zero.)  If we subtract the model [\ion{N}{2}] $\lambda$6583 component from the observed [\ion{N}{2}] $\lambda$5755 map, we obtain the \lq\lq [\ion{N}{2}] $\lambda$5755 excess" map (second row, left).  The map of the excess [\ion{N}{2}] $\lambda$5755 emission is very similar to the \ion{O}{2} $\lambda\lambda$4639,4649 PV diagram (third row, left; see Fig. \ref{fig_pv_recombination} for more examples of PV diagrams of recombination lines), which suggests that this excess emission is indeed due to recombination \citep[c.f.][]{liuetal2000}.  

We can use this decomposition to compute the [\ion{N}{2}] temperature, approximately correcting for the recombination contribution to [\ion{N}{2}] $\lambda$5755.  At bottom right in Fig. \ref{fig_pv_n2}, we present the [\ion{N}{2}] temperature map computed after subtracting the model \ion{O}{2} $\lambda\lambda$4639,4649 emission component from the observed [\ion{N}{2}] $\lambda$5755 PV diagram.  This [\ion{N}{2}] temperature map is now similar to the [\ion{O}{3}] temperature map in Fig. \ref{fig_pv_conditions}, though of much lower signal-to-noise near the central star.  

The agreement of the temperatures calculated from the lines of [\ion{O}{3}], [\ion{Ne}{3}], and [\ion{N}{2}] (corrected for the recombination contribution to [\ion{N}{2}] $\lambda$5755) implies that the quality of the atomic data for these collisional transitions is reasonable.  Therefore, we conclude that recombination excitation of the [\ion{N}{2}] $\lambda$5755 line \citep{nussbaumerstorey1984} is the most plausible explanation for the difference in the [\ion{O}{3}] and [\ion{N}{2}] temperature maps, as has been argued previously for NGC 7009 and other planetary nebulae \citep[][]{liuetal2000, tsamisetal2003, wessonetal2005}.  

In theory, recombination should excite both the auroral and nebular lines of N$^+$ and O$^+$ \citep{nussbaumerstorey1984}.  However, the recombination contribution to the intensities for auroral and nebular lines are similar while the collisional contribution to the nebular lines is 1-2
orders of magnitude larger \citep{liuetal2000}.  The difference between the PV diagrams of [\ion{N}{2}] $\lambda$6583 and [\ion{O}{2}] $\lambda$3726 compared to that of \ion{O}{2} $\lambda\lambda$4639,4649 (Figs. \ref{fig_pv_diagrams} and \ref{fig_pv_n2}) illustrate this clearly for NGC 7009.  Therefore, the recombination contribution to the nebular lines can be ignored in NGC7009 (and probably most other PNe).  

The line splitting also differs between [\ion{O}{3}] $\lambda$4363 and [\ion{O}{3}] $\lambda$4959 (Table \ref{tab_wavelengths}).  In this case, however, the PV diagrams have very similar morphologies (Fig. \ref{fig_pv_diagrams}).   
In addition, the [\ion{O}{3}] $\lambda$4363 PV diagram has no components in common with the PV diagram for \ion{O}{3} $\lambda$3707 that results from recombination of O$^{3+}$, as expected since the quantity of O$^{3+}$ available to recombine is very small outside the innermost part of the O$^{2+}$ zone (see Fig. \ref{fig_ionization_state_model}).  (N$^{2+}$ recombination to [\ion{N}{2}] $\lambda$5755 is much more relevant because there is much more N$^{2+}$ than N$^+$; Fig. \ref{fig_ionization_state_model}.)
 
The O$^{2+}$ ion is the most efficient coolant available in most planetary nebulae.  So, it is natural that the temperature rise where O$^{2+}$ is not the dominant ionization state of oxygen, as occurs in the He$^{2+}$ zone where the fraction of O$^{3+}$ begins to increase.  It is therefore also natural that the [\ion{O}{3}] $\lambda$4363 emission is slightly biased to the inner regions of the O$^{2+}$ zone.  Similarly, in the SW cap, there is a slight increase in the [\ion{N}{2}] (and [\ion{O}{3}]) temperature where the [\ion{O}{1}] $\lambda$6300 emission peaks (Fig. \ref{fig_pv_conditions}), presumably indicating the position of a transition zone and the loss of N$^+$ as an important coolant.  

\subsection{\ion{C}{2} $\lambda$4267}

\begin{figure*}
\begin{center}
\includegraphics[width=2.15\columnwidth]{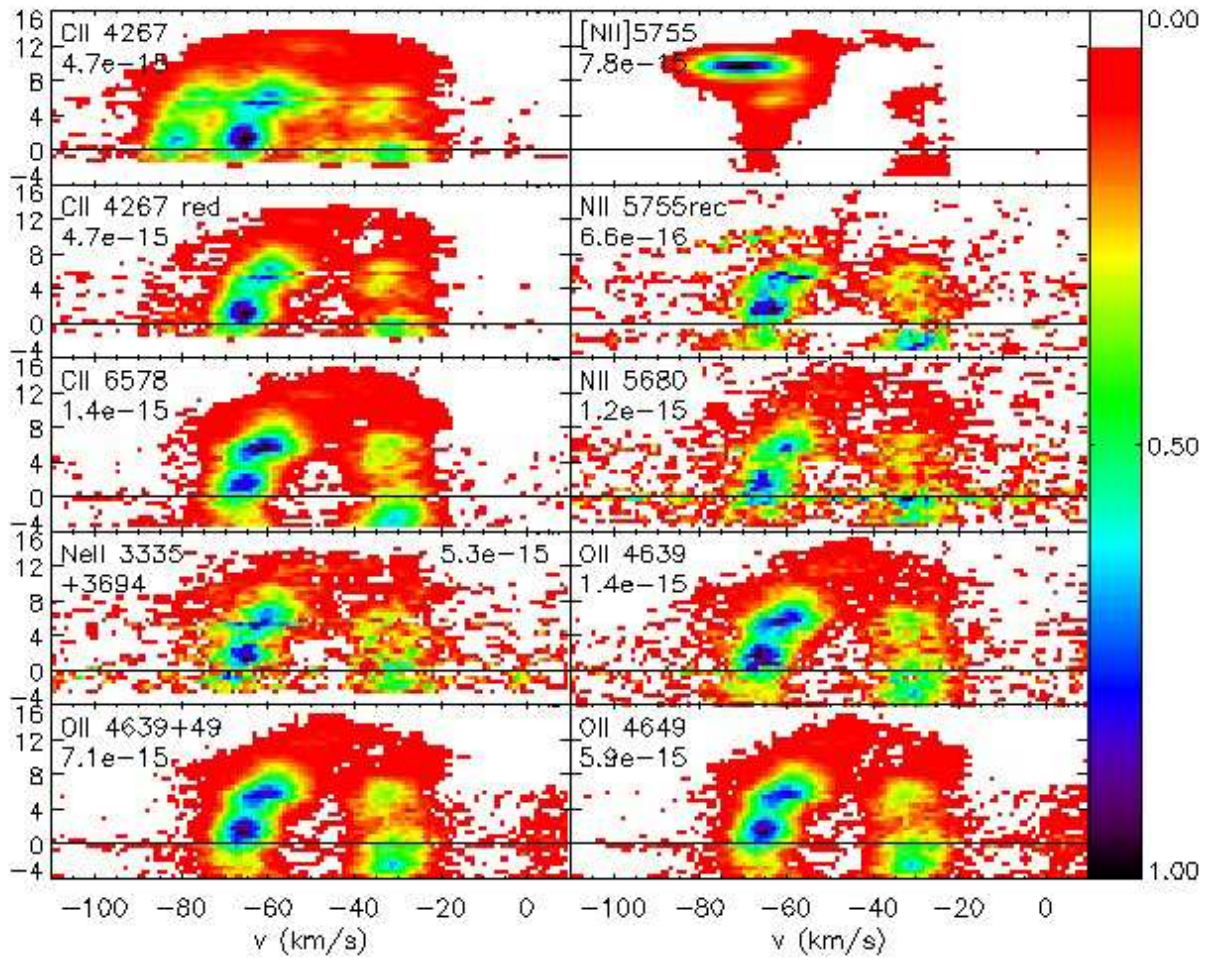}  
\end{center}
\caption{
The PV diagrams in \ion{C}{2}, \ion{N}{2}, \ion{O}{2}, and \ion{Ne}{2} recombination lines are strikingly similar.  Excepting \ion{C}{2} 6578, even the relative intensities of the different components are the same: the blue-shifted emission is always brighter than the red-shifted emission and the components closest to the central star are brighter than the more distant components.  The PV diagrams in the second row are derived from those in the first row (see text).  \ion{C}{2} 6578 and \ion{N}{2} 5680 can be excited via recombination or by indirect continuum fluorescence.  The white level extends to 4\% of the maximum intensity.  See Fig. \ref{fig_pv_diagrams} for further details.  In the panel for \ion{Ne}{2} $3335 + 3694$, the maximum intensity is shown at upper right for clarity.
}
\label{fig_pv_recombination}
\end{figure*}

In Fig. \ref{fig_pv_recombination}, we present more PV diagrams for recombination lines of \ion{C}{2}, \ion{N}{2}, \ion{O}{2}, and \ion{Ne}{2}.  Their similarity is notable in both the positions and velocities of the four main emission components, the two peaks in both the red- and blue-shifted components.  With the exception of the \ion{C}{2} $\lambda$6578 line, the relative intensities of these components are maintained.  Note that, in addition to recombination, fluorescence may contribute indirectly to the excitation of \ion{C}{2} $\lambda$6578 and \ion{N}{2} $\lambda$5680, i.e., fluorescence can excite levels that can then decay to the upper level of these transitions \citep{grandi1976, sharpeeetal2004}.  The decomposition of the [\ion{N}{2}] $\lambda$5755 line has already been discussed (\S \ref{section_n2_5755}).  

The \ion{C}{2} $\lambda$4267 PV diagram is surprising (Fig. \ref{fig_pv_recombination}).  It appears to be composed of only two emission components, not the three expected, i.e.,  (NIST; Atomic Line List v. 2.05b16/P. van Hoof\footnote{http://www.pa.uky.edu/$^{\sim}$peter/newpage/}).  Presumably, we do not detect the $2\mathrm s^2\,4\mathrm f\, ^2\mathrm F^{\circ}_{5/2}$ to $2\mathrm s^2\,3\mathrm d\, ^2\mathrm D_{5/2}$ transition, the weakest and reddest component.  Also, the separation between the two components we observe, the $2\mathrm s^2\,4\mathrm f\, ^2\mathrm F^{\circ}_{7/2}$ to $2\mathrm s^2\,3\mathrm d\, ^2\mathrm D_{5/2}$ and $2\mathrm s^2\,4\mathrm f\, ^2\mathrm F^{\circ}_{5/2}$ to $2\mathrm s^2\,3\mathrm d\, ^2\mathrm D_{3/2}$ transitions, is 0.222\AA\ (15.6\,km/s) rather than the 0.182\AA\ or 0.260\AA\ expected (NIST; Atomic Line List v. 2.05b16/P. van Hoof$^4$).  

We have isolated the red emission component (Fig. \ref{fig_pv_recombination}, second row, left panel), nominally at 4267.183/4267.261\AA, by modeling the \ion{C}{2} $\lambda$4267 PV diagram as a linear superposition of two \ion{O}{2} $\lambda\lambda$4639,4649 PV diagrams that are shifted in velocity.  The PV diagram for the red component is obtained by subtracting the blue component from the original \ion{C}{2} 4267 PV diagram.  (As before, the intercept of the fit was forced through the origin.)  

\subsection{\ion{N}{2} fluorescence lines}\label{section_n2_fluorescence}

Several \ion{N}{2} multiplets may be excited by fluorescence from starlight or \ion{He}{1} $\lambda$508.64 \citep[e.g.,][]{grandi1976, sharpeeetal2004}.  As already noted, \ion{N}{2} $\lambda$5680 and other lines of its multiplet (V3) can result from the de-excitation of the lines from multiplet V30 (\ion{N}{2} $\lambda\lambda$3829.80,3842.19,3855.10) that are strongly affected by resonance fluorescence from \ion{He}{1} $\lambda$508.64, as can multiplet V5, of which \ion{N}{2} $\lambda$4630.54 is the strongest component.  Resonance fluorescence by starlight can excite multiplets V20, V21, V24, V28, and V29, of which the strongest lines are \ion{N}{2} $\lambda$4994.37 (V24) and \ion{N}{2} $\lambda\lambda$5931.79,5941.65 (V28).  Only \ion{N}{2} $\lambda\lambda$4631,5932,5942 have sufficient signal to construct (noisy) PV diagrams (as well as \ion{N}{2} $\lambda$5680 and other lines from multiplet V3).  

\begin{figure}
\begin{center}
\includegraphics[width=1.02\columnwidth]{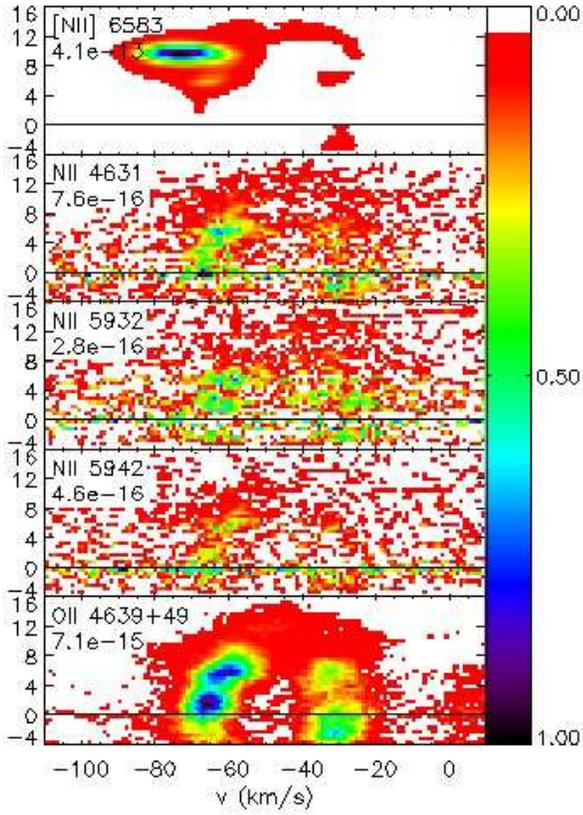}  
\end{center}
\caption{
We compare the collisionally-excited [\ion{N}{2}] $\lambda$6583 line with the \ion{N}{2} $\lambda\lambda$4631,5932,5942 lines that can be excited via fluorescence and the \ion{O}{2} $\lambda\lambda$4639,4649 recombination lines.  In spite of the low signal-to-noise, the \ion{N}{2} $\lambda\lambda$4631,5932,5942 PV diagrams are more similar to that of \ion{O}{2} $\lambda\lambda$4639,4649, indicating that this process likely dominates their production in NGC 7009.  The white level extends to 3\% of the maximum intensity.  See Fig. \ref{fig_pv_diagrams} for further details.
}
\label{fig_pv_n2_fluorescence}
\end{figure}

If these lines are excited purely by recombination, their PV diagrams should appear similar to other recombination lines, such as \ion{O}{2} $\lambda\lambda$4639,4649.  If they result purely from fluorescence, their PV diagrams should be more similar to the PV diagram of [\ion{N}{2}] $\lambda$6583.  Fig. \ref{fig_pv_n2_fluorescence} presents the PV diagrams for the strongest of the \ion{N}{2} fluorescence lines in our data, \ion{N}{2} $\lambda\lambda$4631,5932,5942 as well as [\ion{N}{2}] $\lambda$6583 and \ion{O}{2} $\lambda\lambda$4639,4649.  (\ion{N}{2} $\lambda$5680 appears in Fig. \ref{fig_pv_recombination}.)  Like the \ion{N}{2} $\lambda$5680 PV diagram, those for \ion{N}{2} $\lambda\lambda$4631,5932,5942, though noisy, are more similar to the PV diagrams of recombination lines than to that for [\ion{N}{2}] $\lambda$6583.  We conclude that, in NGC 7009, these lines are excited overwhelmingly via recombination, confirming the opinion of \citet{sharpeeetal2004} that different excitation mechanisms have different effects in different objects.  

\section{Discussion}\label{section_discussion}

\subsection{Modeling the ionization structure}\label{section_ionization_structure_discussion}

\begin{figure*}
\begin{center}
\includegraphics[width=2.1\columnwidth]{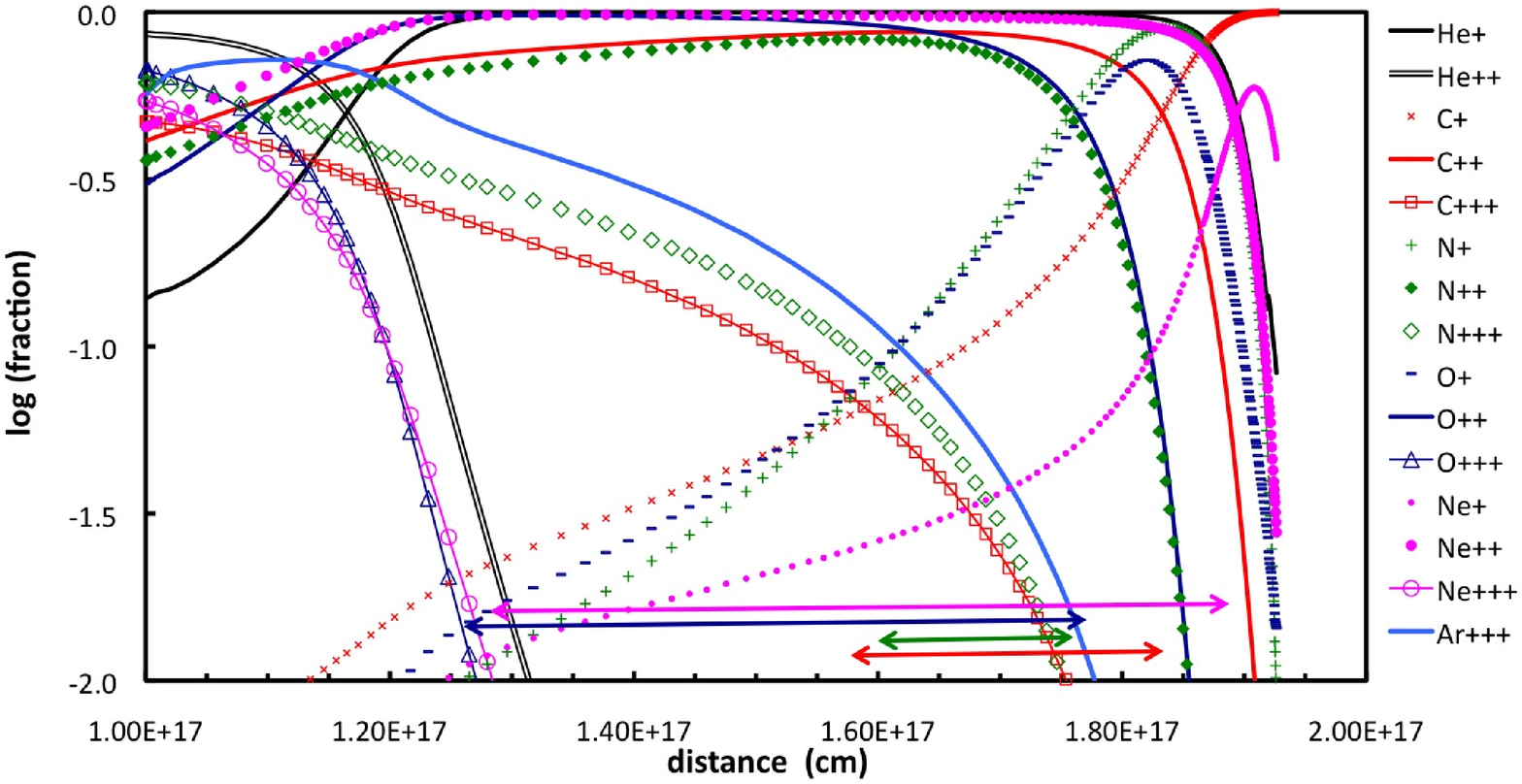}  
\end{center}
\caption{
We plot the fractional ionization in a model nebula as a function of distance from the ionizing source.  This optically-thick model was computed using v08.00 of CLOUDY, last described by \citet{ferlandetal2013}, assuming an inner radius of 10$^{17}$\,cm \citep[][their Fig. 9, top panel]{sabbadinetal2004}, a constant density of 4000\,cm$^{-3}$, a filling factor of unity, a homogeneous chemical composition (CLOUDY's default abundance set for planetary nebulae), and a stellar atmosphere model from \citet[][100,000\,K, $\log g=5$, $\log L_{bol} = 38.6$]{rauch2003}.  
The red, green, blue, and magenta arrows at bottom indicate the zones from which emission should arise from \ion{C}{2}, \ion{N}{2}, \ion{O}{2}, and \ion{Ne}{2}, respectively.
}
\label{fig_ionization_state_model}
\end{figure*}

\begin{figure}
\begin{center}
\includegraphics[width=\columnwidth]{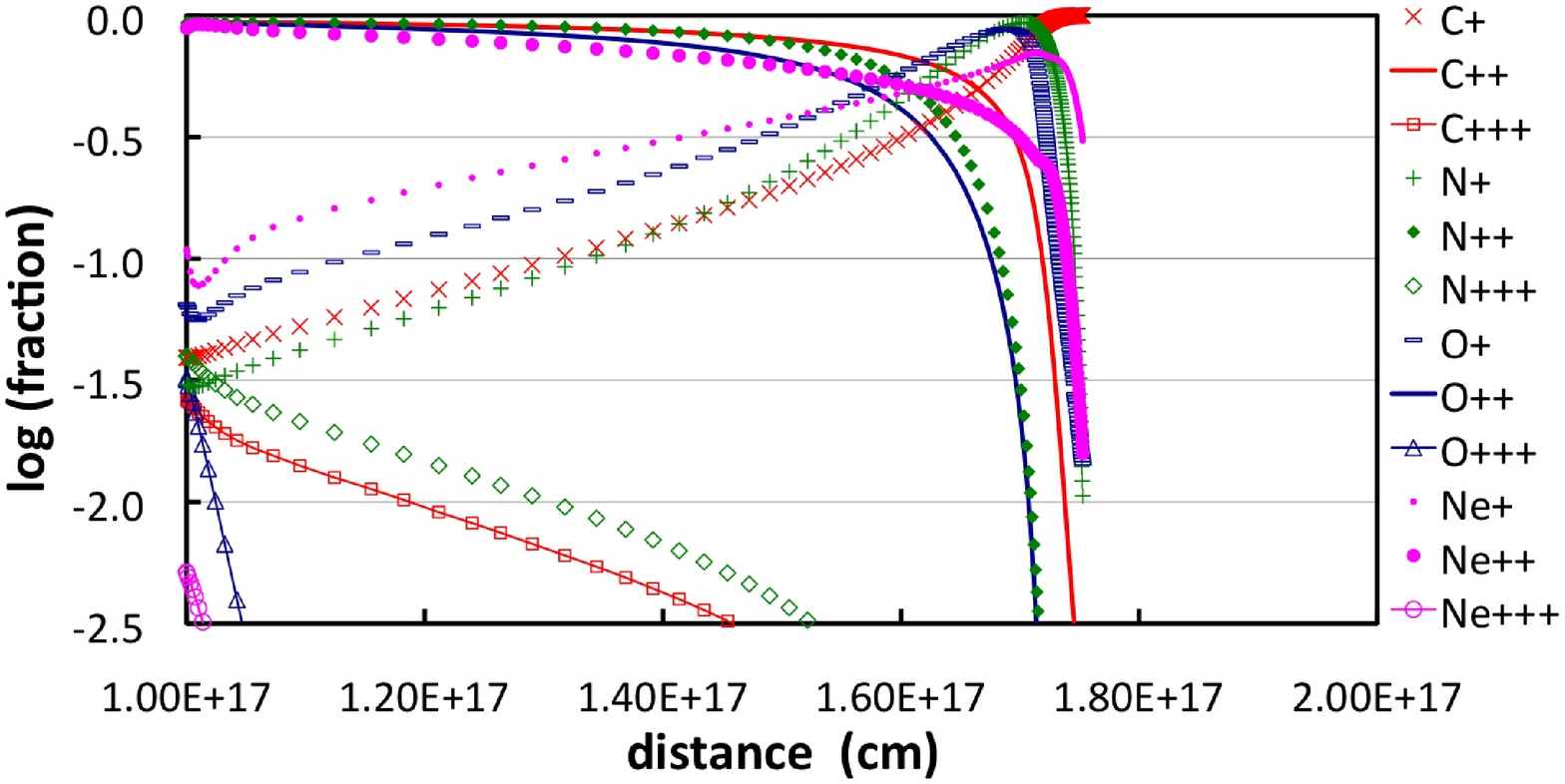}  
\includegraphics[width=\columnwidth]{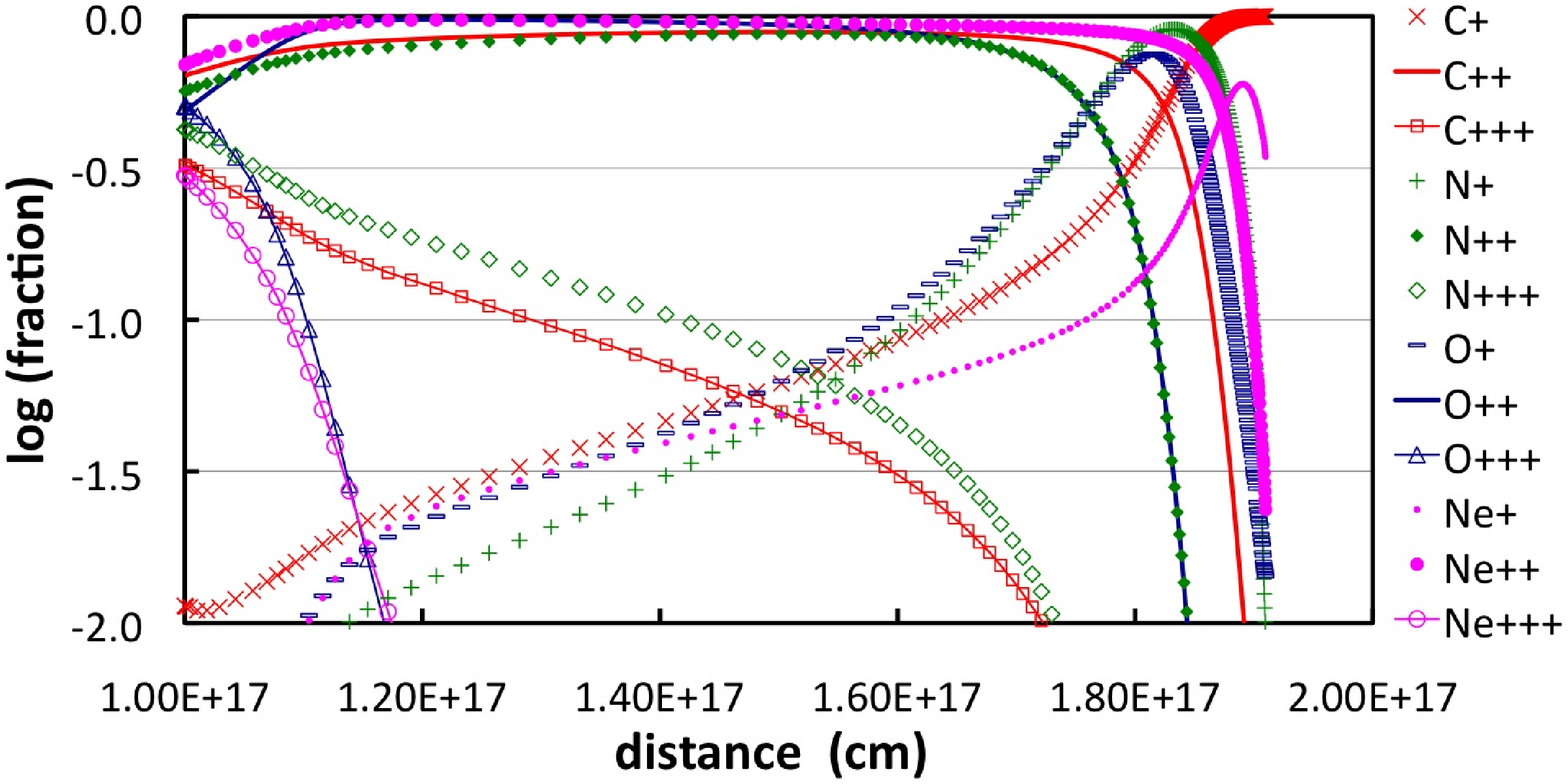}  
\includegraphics[width=\columnwidth]{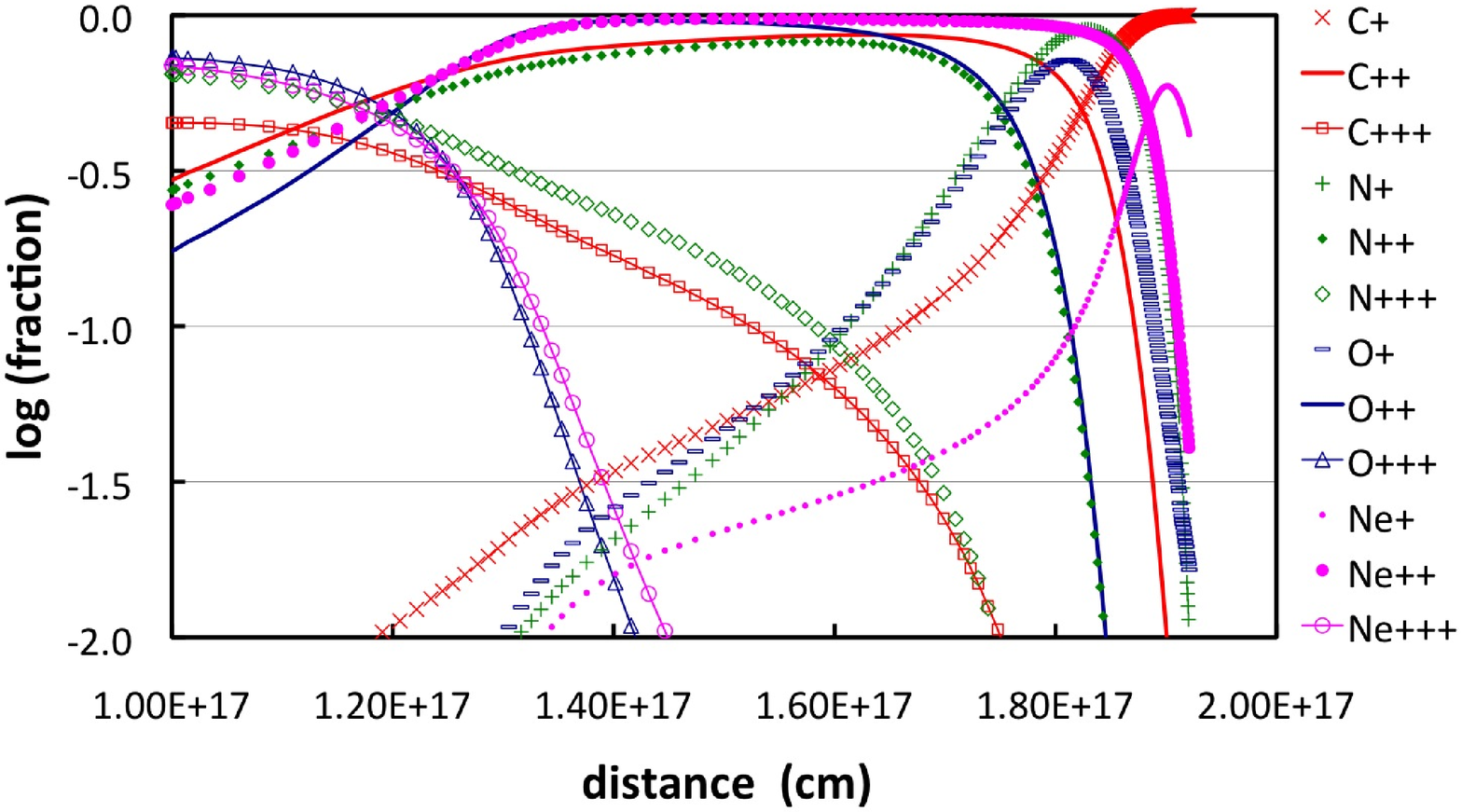}  
\includegraphics[width=\columnwidth]{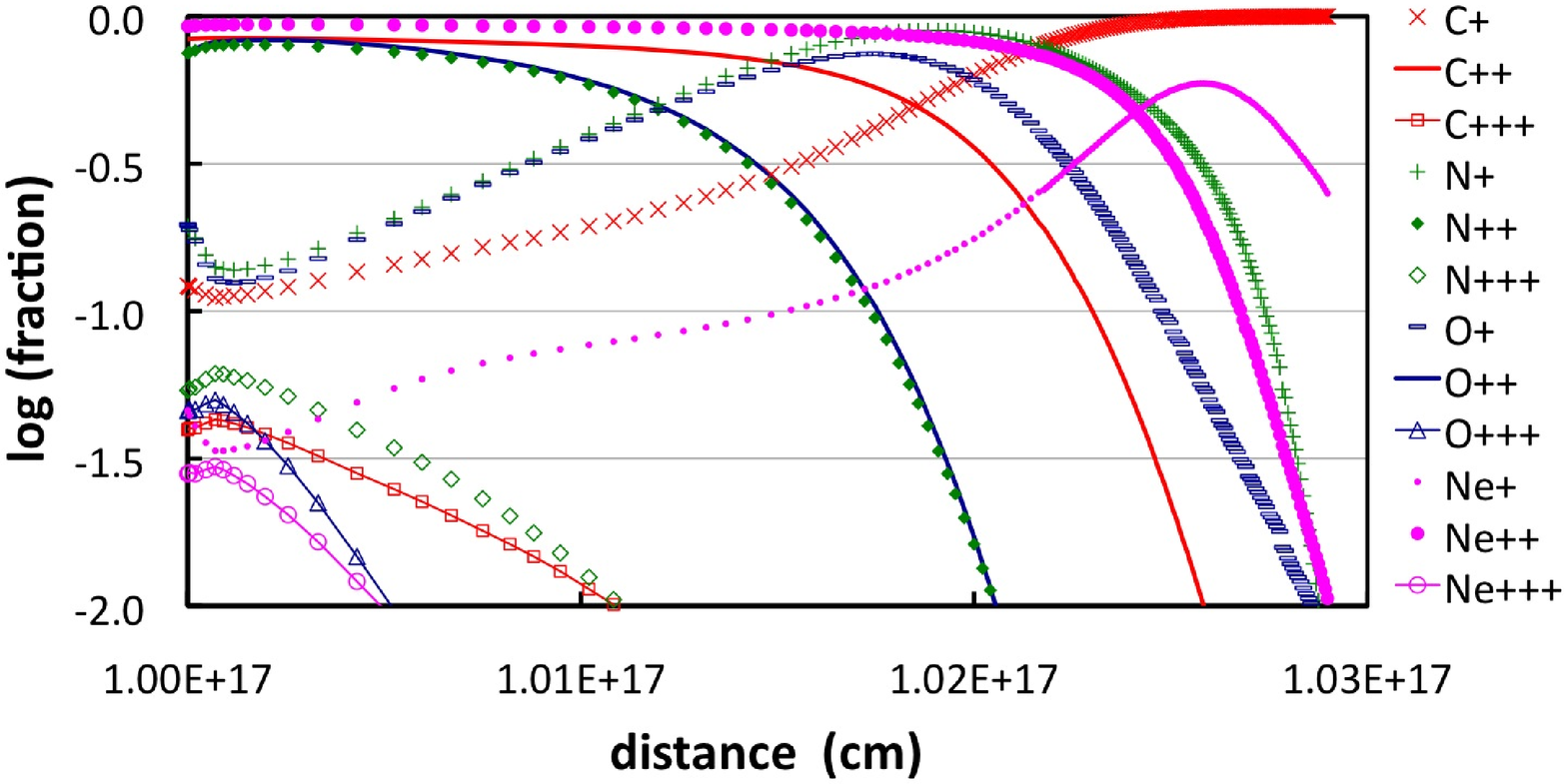}  
\end{center}
\caption{
We present additional models whose parameters are varied from those in Fig. \ref{fig_ionization_state_model}.  In the top three panels, the central stars are modeled as black bodies with temperatures of 60,000\,K (top), 80,000\,K (second), and 120,000\,K (third).  In the bottom panel, the density is increased by a factor of 10 (note the change in horizontal scale).  Otherwise, the parameters are as in Fig. \ref{fig_ionization_state_model}.  For clarity, we omit the He$^{2+}$, He$^+$, and Ar$^{3+}$ ionization fractions.
}
\label{fig_models_grid}
\end{figure}

To provide context for the results of \S \ref{section_ionization_structure}, we present the ionization structure for generic model nebulae in Figs. \ref{fig_ionization_state_model} and \ref{fig_models_grid}.  Both figures present the fractional ionization as a function of distance from the ionizing source.  The figure captions provide details concerning the model parameters.  The models in Fig. \ref{fig_ionization_state_model} and the second panel of Fig. \ref{fig_models_grid} contain the ionizing sources most likely to be similar to the central star in NGC 7009.  The ionization structure of these two models is very similar to that deduced from the kinematics of NGC 7009:  The dominant ions throughout the bulk of the main shell, especially if it is optically thin, are H$^+$, He$^+$, O$^{2+}$, and Ne$^{2+}$.  The $\mathrm O^{3+} \rightarrow \mathrm O^{2+}$ transition zone occurs within the He$^{2+}$ zone and more or less coincides with the Ar$^{3+}$ zone.  

In all cases, the zones where C$^{2+}$, N$^{2+}$, O$^{2+}$, and Ne$^{2+}$ are the dominant ions of these elements coincide rather closely.  The \ion{C}{2}, \ion{N}{2}, \ion{O}{2}, and \ion{Ne}{2} recombination lines should arise primarily where these doubly-ionized ions recombine to the singly ionized state, i.e., where the singly-ionized stage of these elements is the \emph{second} most abundant stage (see the arrows in Fig. \ref{fig_ionization_state_model}).  For the models with a significant He$^{2+}$ zone, as in NGC 7009, the zone where C$^+$ and N$^+$ are the second most abundant stage differs significantly from the zone where O$^+$ and Ne$^+$ are the second most abundant stage, since the C$^{3+}$ and N$^{3+}$ zones extend deeply into the C$^{2+}$ and N$^{2+}$ zones, respectively.    
Therefore, these models predict different kinematics for \ion{C}{2} and \ion{N}{2} lines, in particular a larger velocity splitting, compared to \ion{O}{2} and \ion{Ne}{2} lines.  In contrast, we observe similar kinematics for all of these lines in NGC 7009.

In reality, this problem may not be so acute, because NGC 7009's main shell is optically thin to ionizing radiation.  As a result, the models in Fig. \ref{fig_ionization_state_model} and the top three panels of Fig. \ref{fig_models_grid} should be truncated prior to the $\mathrm H^+ \rightarrow \mathrm H^{\circ}$ transition zone.  Depending upon where the models are truncated, the \ion{C}{2} and \ion{N}{2} lines should be significantly weaker than expected due to ionization equilibrium, but would allow the zones emitting the \ion{C}{2}, \ion{N}{2}, \ion{O}{2}, and \ion{Ne}{2} recombination lines to coincide more closely, perhaps explaining their similar kinematics.  

On the other hand, truncating the models as described is problematic as regards the abundances derived from the \ion{C}{2}, \ion{N}{2}, \ion{O}{2}, and \ion{Ne}{2} lines.  From the truncated models, lower C$^{2+}$ and N$^{2+}$ abundances would be expected as compared to the O$^{2+}$ and Ne$^{2+}$ abundances, since a large fraction of the C$^{2+}$ and N$^{2+}$ zones is excluded.  Yet, the over-abundance of C$^{2+}$, N$^{2+}$, O$^{2+}$, and Ne$^{2+}$ derived from recombination lines in NGC 7009 are similar \citep{liuetal1995, luoliu2003, fangliu2012}.    

So, the models in Fig. \ref{fig_ionization_state_model} and the top three panels of Fig. \ref{fig_models_grid} are not simultaneously compatible with the kinematics and chemical abundances from recombination lines in NGC 7009.  To be compatible with the kinematics we observe, the models should be truncated.  If they are truncated, the models should be incompatible with the ionic over-abundances derived from the \ion{C}{2}, \ion{N}{2}, \ion{O}{2}, and \ion{Ne}{2} lines \citep{liuetal1995, luoliu2003, fangliu2012}.  

\subsection{Location of the ions emitting \ion{C}{2}, \ion{N}{2}, \ion{O}{2}, and \ion{Ne}{2} lines}\label{section_location_recombination}

The O$^{2+}$ ion is responsible for four PV diagrams in Fig. \ref{fig_pv_diagrams}, the collisionally-excited [\ion{O}{3}] $\lambda\lambda$4363,4959 lines, the \ion{O}{3} $\lambda$3444 Bowen fluorescence line, and the \ion{O}{2} $\lambda\lambda$4639,4649 recombination lines (also Table \ref{tab_wavelengths}).  The structure of the PV diagram of the \ion{O}{3} $\lambda$3444 Bowen fluorescence line is notably different from that of the [\ion{O}{3}] $\lambda\lambda$4363,4959 or \ion{O}{2} $\lambda\lambda$4639,4649, but resembles that for \ion{He}{2} $\lambda$4686 in that the \lq\lq loop" at the top/SW of the diagram is replaced by a broad, detached kinematic component that shows little spatial structure.  

The Bowen and collisionally-excited lines are expected to differ since the spatial volumes in which they are excited differ.  The PV diagrams differ in the sense expected, since the Bowen lines should arise in the innermost part of the O$^{2+}$ zone that overlaps the He$^{2+}$ zone.  Indeed, both the structure of the PV diagram of the \ion{O}{3} $\lambda$3444 line and its line splitting are similar to those of the \ion{He}{2} $\lambda$4686 line, as expected.

The structure of the PV diagrams for the [\ion{O}{3}] $\lambda\lambda$4363,4959 collisional lines are very similar.  However, their line splitting differs (Table \ref{tab_wavelengths}).  This difference was addressed in \S\S \ref{section_ionization_structure} and \ref{section_physical_conditions} and can be understood in terms of the ionization and temperature structure of the nebula (\S \ref{section_ionization_structure_discussion}).  

That the PV diagrams for [\ion{O}{3}] $\lambda$4959 and \ion{O}{2} $\lambda\lambda$4639,4649 differ is very important.  In the absence of strong temperature gradients, ionization equilibrium in a chemically-homogeneous medium should impose similar position and velocity distributions on the collisional and recombination emission from O$^{2+}$ (\S\S \ref{section_ionization_structure} and \ref{section_ionization_structure_discussion}; Figs. \ref{fig_ionization_state_model} and \ref{fig_models_grid}).  Since the emissivity of recombination and collisionally-excited lines depend upon the temperature in opposite ways, the former decreasing and the latter increasing as the temperature increases, a temperature gradient could enhance the \ion{O}{2} $\lambda\lambda$4639,4649 recombination emission from the coolest regions in a chemically-uniform plasma.  

The top panel in Fig. \ref{fig_pv_o2} presents the PV diagram of the ratio of the emission in \ion{O}{2} $\lambda\lambda$4639,4649 relative to that in the [\ion{O}{3}] $\lambda$4959 line.  If the \ion{O}{2} $\lambda\lambda$4639,4649 emission arose only from recombination of O$^{2+}$ ions throughout the zone that emits [\ion{O}{3}] $\lambda$4959, the ratio should be constant in the outer part of the nebula and decrease in the inner part of the O$^{2+}$ zone where the temperature increases (bottom panel in Fig. \ref{fig_pv_o2}; also Fig. \ref{fig_pv_conditions}).  While a  gradient is seen, it is in the wrong sense, for it increases towards the innermost part of the zone emitting [\ion{O}{3}] $\lambda$4959 where the temperature increases, rather than decreasing as expected.  

The second and third panels of Fig. \ref{fig_pv_o2} present the PV diagrams of the ratios of \ion{O}{3} $\lambda$3444 and [\ion{Ar}{4}] $\lambda$4740 relative to [\ion{O}{3}] $\lambda$4959.  These two line ratios vary in a similar way as the ratio of \ion{O}{2} $\lambda\lambda$4629,4649 to [\ion{O}{3}] $\lambda$4959.  However, both the Ar$^{3+}$ zone and the \ion{O}{3} $\lambda$3444 Bowen fluorescence are biased to the innermost part of the O$^{2+}$ zone within the He$^{2+}$ zone (Fig. \ref{fig_ionization_state_model}), so their line ratios relative to [\ion{O}{3}] $\lambda$4959 vary as expected and confirm that the excess \ion{O}{2} $\lambda\lambda$4629,4649 in the inner part of the O$^{2+}$ zone is anomalous.    

The structure of the top two panels in Fig. \ref{fig_pv_o2} differ in an important way.  Clearly, the \ion{O}{3} $\lambda$3444 Bowen fluorescence (second panel) is much more restricted to the inner part (in velocity) of the O$^{2+}$ zone than the emission from \ion{O}{2} $\lambda\lambda$4639,4649 (top panel).  This is especially apparent considering the velocity-space range occupied by the SW cap (Figs. \ref{fig_pv_explanation} and \ref{fig_pv_diagrams}).  Thus, there is emission from \ion{O}{2} $\lambda\lambda$4639,4649 from the entire zone from which the [\ion{O}{3}] $\lambda$4959 emission arises, but this is not the case for the emission from \ion{O}{3} $\lambda$3444.  Therefore, the entire volume emitting in the [\ion{O}{3}] $\lambda$4959 forbidden line emits some \ion{O}{2} $\lambda\lambda$4639,4649 recombination emission, as expected.  What is unexpected is the additional emission from \ion{O}{2} $\lambda\lambda$4639,4649 in the inner part of the O$^{2+}$ zone.  Given what is known of the temperature structure in NGC 7009 and the expectations of ionization equilibrium, the \ion{O}{2} $\lambda\lambda$4639,4649 emission clearly does not trace the emission from [\ion{O}{3}] $\lambda$4959 in the inner part of the O$^{2+}$ zone.  Thus, there appear to be two emission components for the \ion{O}{2} $\lambda\lambda$4639,4649 recombination lines in NGC 7009: one probably follows the [\ion{O}{3}] $\lambda$4959 emission throughout the O$^{2+}$ zone as expected while the second is located near the $\mathrm O^{3+}\rightarrow \mathrm O^{2+}$ transition zone.  

Our data (Table \ref{tab_wavelengths}, Figs. \ref{fig_pv_diagrams} and \ref{fig_pv_recombination}) lead us to identical conclusions concerning the \ion{Ne}{2} recombination lines and the collisionally-excited [\ion{Ne}{3}] lines in NGC 7009.  Although we lack CELs to trace the C$^{2+}$ and N$^{2+}$ zones, the models presented in \S \ref{section_ionization_structure_discussion} imply that the same would also be true for the \ion{C}{2} and \ion{N}{2} recombination lines.

There are few studies of the kinematics of the \ion{C}{2}, \ion{N}{2}, \ion{O}{2}, and \ion{Ne}{2} lines.  \citet[][using an ADC]{otsukaetal2010} also found smaller expansion velocities from \ion{O}{2} that were discrepant with the expectations of ionization equilibrium in the planetary nebula BoBn 1 ($\mathrm{ADF}\sim 2-3$).  On the other hand, the kinematics of the \ion{C}{2}, \ion{N}{2}, and \ion{Ne}{2} lines was not discrepant.  For DdDm 1 ($\mathrm{ADF}\sim 12$), \citet{otsukaetal2009} find that the kinematics of the \ion{N}{2} and \ion{O}{2} lines is compatible with the ionization structure derived from CELs.  The results of \citet[][IC 418]{sharpeeetal2004} and \citet[][NGC 6153 and NGC 7009]{barlowetal2006} are more difficult to interpret, since they were made without an ADC and on the bright rims of these objects, which cannot guarantee co-spatial observations at different wavelengths \citep[e.g.,][]{fillipenko1982} and can confuse ionization structure with projection effects, respectively.  Only for NGC 6153 ($\mathrm{ADF}\sim 9$) does the expansion velocity of \ion{O}{2} $\lambda$4649 appear to be smaller than that for [\ion{O}{3}] $\lambda$5007 \citep{barlowetal2006}.

On the other hand, differences between the spatial distributions of CELs and ORLs of carbon and oxygen are well-known.   
\citet{liuetal2000}, \citet{garnettdinerstein2001}, \citet{luoliu2003}, and \citet{tsamisetal2008} find that the spatial profile of the \ion{O}{2} recombination lines is more centrally-concentrated than that for the forbidden [\ion{O}{3}] emission in NGC 6153, NGC 7009, NGC 6720, and (marginally) NGC 5882.   \citet{barker1982,barker1991} noted a similar effect for \ion{C}{2} $\lambda$4267 compared to \ion{C}{3}] $\lambda\lambda$1906,1909 in NGC 6720 and NGC 2392.  Supposing spherical or cylindrical symmetry, these differences in spatial profiles are compatible with the differences in kinematics we observe in NGC 7009.  

We may therefore infer that the distributions of recombination and collisionally-excited lines through NGC 7009 differ, both \emph{projected in the plane of the sky} and \emph{along the line of sight}.  Physically, one component of the \ion{O}{2} $\lambda\lambda$4639,4649 recombination emission arises closer to the central star than does the majority of the collisionally-excited emission, i.e., the kinematics of this component of the \ion{O}{2} $\lambda\lambda$4639,4649 recombination lines are like those of higher stages of ionization in NGC 7009.  Given the similarity of their kinematics, the same is true for the \ion{C}{2}, \ion{N}{2}, and \ion{Ne}{2} emission.  Therefore, there is presumably the same additional emission component for all of these recombination lines and the volume from which it arises is \emph{not} the same volume as that from which collisionally-excited lines from the parent ions are emitted, as traced by [\ion{O}{3}] $\lambda$4959 and [\ion{Ne}{3}] $\lambda$3869.  While we cannot say whether these volumes are mutually exclusive (\S \ref{section_cold_clumps}), they are clearly well-mixed in the inner part of the O$^{2+}$ zone (Fig. \ref{fig_pv_o2}).  

\subsection{Cold clumps?}
\label{section_cold_clumps}

To detect cold clumps kinematically requires sufficient spectral resolution to at least resolve the thermal widths of the normal nebular plasma (emitting in CELs, $\mathrm{resolution} > 10^5$) and demonstrate that the ORLs have narrower line widths.  Our spectral resolution prevents this.  However, even very high spectral resolution could be foiled if there are multiple clumps of the cold and normal plasmas at different velocities along the line of sight.  

A less direct method for inferring the presence of two plasma components, one emitting the ORLs, the other the CELs, is to search for strong, local enhancements in the ratio of ORL to CEL emission.  This can be done easily with the ratio of two PV diagrams.  Again, very high spectral resolution is advantageous, especially if there are multiple clumps of the different plasma components along the line of sight.  Our observations (Fig. \ref{fig_pv_o2}) do not reveal spectrally resolved enhancements of \ion{O}{2} $\lambda\lambda$4639,4649 relative to [\ion{O}{3}] $\lambda$4959, but a clear gradient, as already discussed.  Thus, either (1) our spectral resolution smears both the ORL and CEL emission too much, (2) there are multiple clumps of the ORL- and CEL-emitting plasma components along all of our lines of sight through NGC 7009, or (3) the ORL-emitting plasma is widely distributed within the normal nebular plasma.  

The fourth panel in Fig. \ref{fig_pv_o2} clearly shows that the \ion{He}{1} $\lambda$4922 emission does not have the spatial and velocity distribution of the \ion{O}{2} $\lambda\lambda$4639,4649 emission.  In spite of the poorer velocity resolution (the [\ion{O}{3}] $\lambda$4959 PV diagram was broadened to match the thermal width of the \ion{He}{1} $\lambda$4922 line), it is clear that the \ion{He}{1} $\lambda$4922 line emission closely follows that from [\ion{O}{3}] $\lambda$4959, as expected 
(Figs. \ref{fig_ionization_state_model} and \ref{fig_models_grid}).  Clearly, the plasma component emitting the \ion{C}{2}, \ion{N}{2}, \ion{O}{2}, and \ion{Ne}{2} lines in the inner part of the O$^{2+}$ zone does not emit particularly strongly in \ion{He}{1} $\lambda$4922.  Hence, our observations can only support a metal-rich plasma component emitting the \ion{C}{2}, \ion{N}{2}, \ion{O}{2}, and \ion{Ne}{2} lines in the inner part of the O$^{2+}$ zone.  

So, we find no clear evidence of chemically-distinct plasma components, such as cold, hydrogen-deficient clumps in NGC 7009 \citep[e.g.,][]{liuetal2000}, but we cannot rule them out.  In principle, they could give rise to the observed recombination lines at the locations and velocities observed (or anywhere, since there is no accepted explanation for their origin).  If they were considerably more dense than the \lq\lq normal" nebular plasma (allowing pressure equilibrium), they would naturally be of lower ionization degree, which would resolve the problem of coincident emission from \ion{C}{2}, \ion{N}{2}, \ion{O}{2}, and \ion{Ne}{2} (\S \ref{section_ionization_structure_discussion}).  The dense model in the bottom panel of Fig. \ref{fig_models_grid} would not be spatially resolved at the distance of NGC 7009.

The hypothesis of dense clumps, however, is incompatible with the physical conditions derived from the \ion{N}{2} and \ion{O}{2} lines in NGC 7009 by \citet{fangliu2011} and \citet{fangliu2012}.  They find that both the temperature and density derived for these lines is \emph{lower} than those derived for CELs.  If true, the volume emitting these lines is not in pressure equilibrium with the volume emitting the CELs.  We note that the atomic data that these authors used for \ion{O}{2} has not been published.

\subsection{Other options?}

The [\ion{O}{3}] temperature map in Fig. \ref{fig_pv_conditions} clearly shows that measurable, large-scale temperature gradients exist within NGC 7009.  Small-scale temperature fluctuations may also exist, but our spectral resolution limits our ability to measure their true amplitude.  However, it is not evident 
that these temperature fluctuations could be responsible for the enhanced \ion{O}{2} $\lambda\lambda$4639,4649 emission.  Our observations, like those of \citet[][]{tsamisetal2008}, find that the \ion{C}{2}, \ion{N}{2}/[\ion{N}{2}], \ion{O}{2}, and \ion{Ne}{2} recombination emission coincides with regions of higher electron temperature, not lower temperatures, contrary to expectations if temperature fluctuations are responsible for enhancing the recombination lines. 

The transition zones of triply to doubly ionized C, N, O, and Ne coincide with the region from which recombination lines of \ion{C}{2}, \ion{N}{2}, \ion{O}{2}, and \ion{Ne}{2} originate in NGC 7009.  Might the resulting doubly ionized ions somehow recombine efficiently to the singly ionized state?  Even if this were feasible in NGC 7009, it could not be a general process because other planetary nebulae with large ADFs, such as DdDm 1 or NGC 40 \citep{liuetal2004a, wessonetal2005, otsukaetal2009}, do not possess triply ionized zones of C, N, O, and Ne.  

Physical processes that affect recombination throughout the O$^{2+}$ zone do not appear promising as an explanation for the kinematics of the \ion{C}{2}, \ion{N}{2}, \ion{O}{2}, and \ion{Ne}{2} lines in NGC 7009.  Although dielectronic recombination could strongly enhance recombination lines where the temperature increases, as we observe, we would require similar enhancements for four ions with different electronic structures, which seems improbable.  Fluorescence would require efficiently exciting minority ions (\S \ref{section_ionization_structure}) and at least the \ion{N}{2} lines in NGC 7009 do not appear to be excited primarily by fluorescence (\S \ref{section_n2_fluorescence}).  \citet{rodriguezgarciarojas2010} and \citet{pradhanetal2011} suggest that recombination coefficients  could explain the differences between ORL and CEL abundances.  While this might address a general effect, such as the typical $\mathrm{ADF}\sim 2$, it cannot explain our kinematics since it should enhance recombination throughout the region where doubly ionized C, N, O, and Ne are present and, like dielectronic recombination, would have to change in a similar way for four different ions.  Likewise, the possibility of departures of the electron energy distribution from a Maxwellian distribution should affect the entire zone where doubly ionized C, N, O, and Ne are present \citep{nichollsetal2012}.  Again, such an effect might explain a general discrepancy (e.g., $\mathrm{ADF}\sim 2$), but it would not produce the component of the \ion{C}{2}, \ion{N}{2}, \ion{O}{2}, and \ion{Ne}{2} emission we observe at the  velocity of the $\mathrm O^{3+} \rightarrow \mathrm O^{2+}$ transition zone in NGC 7009.

\subsection{The abundance discrepancy}\label{section_abundance_discrepancy}

Given that there appear to be at least two \ion{O}{2} emission components in NGC 7009 and that one does not arise from the same volume as the [\ion{O}{3}] emission (\S\S \ref{section_location_recombination} and \ref{section_cold_clumps}), it is clear that care must be taken to compare and interpret the chemical abundances derived from each type of line.  In particular, it is indispensable to know the physical conditions under which each type of line emits.  Thus far, this type of analysis is not common, but \citet{liuetal2006} have undertaken such an analysis for the planetary nebula Hf 2-2.  Under the usual assumptions, Hf 2-2 has an ADF of 72 (the highest known), but, when appropriate temperatures and densities are assumed for the \ion{O}{2}, [\ion{O}{3}], and Balmer lines, comparable masses of O$^{2+}$ are found from CELs and ORLs.  Presumably, in other objects with smaller ADFs, the mass of O$^{2+}$ implied by the \ion{O}{2} lines is less.   

Since one component of the \ion{O}{2} emission arises from a different volume from that giving rise to the [\ion{O}{3}] emission in NGC 7009, both must be measured to obtain the oxygen abundance for these plasma components.  Since the component responsible for the \ion{O}{2} lines also appears to give rise to the \ion{C}{2}, \ion{N}{2}, and \ion{Ne}{2} lines, the same is presumably true for the carbon, nitrogen, and neon abundances.  If the \ion{O}{2} and [\ion{O}{3}] lines arise from different plasma components, the meaning of the abundance \lq\lq discrepancy" changes:  It is no longer that the collisional and recombination lines imply abundances that disagree, but that they probe the chemical abundances in different plasma components and that the different plasma components have different chemical compositions.  In NGC 7009, it is not a matter that one abundance is correct and the other wrong, as is commonly argued, but that they measure the oxygen abundance in different volumes within the object.  

If there are multiple plasma components, it is important to understand the origin of each component in order to interpret the abundances derived for each of them.  The standard analysis of nebular gas assumes a common origin for all of the material, the progenitor star in the case of planetary nebulae.  This standard analysis will be reasonable in many cases, but not all.  When there are multiple plasma components, it is not necessarily obvious that all have a common origin, especially when even their coexistence is not understood.  Furthermore, depending upon the astrophysical interest in the chemical abundances, not all of the plasma components may be relevant.  For example, in studies of the composition of the ejecta of the progenitor star, it is vital to assure that only those components that were part of the progenitor star are considered.
\citet{tsamisetal2011}, \citet{tsamiswalsh2011} and \citet{mesadelgadoetal2012} provide an enlightening study, though for a different context (a proplyd in Orion).  

In planetary nebulae, there are multiple ways of incorporating external material whose origin is not the progenitor star.  A simple example is the sweeping up of interstellar material during the evolution of the nebular shell \citep[e.g.,][]{villaveretal2002}.  Although this mechanism could be relevant for relatively evolved objects when the shell is optically-thin and the analysis includes swept up interstellar material, it cannot explain the kinematics we find for the \ion{C}{2}, \ion{N}{2}, \ion{O}{2}, and \ion{Ne}{2} lines in NGC 7009.  The dust discs found around the central stars of some planetary nebulae could be a more relevant example of external material \citep{suetal2007, chuetal2011, bilikovaetal2012}, which \citet{henneystasinska2010} find may be responsible for the chemical abundances in the surrounding nebulae under some circumstances.  In this scenario, the kinematics of the ORLs need not coincide with those expected from ionization equilibrium, as the sublimating material will be at least partially decoupled from the nebular shell, and so could be compatible with our observations in NGC 7009.  In both cases, the external material is relevant to determine the composition of the nebular plasma, but not that of the matter ejected by the progenitor star.  

The foregoing has far-reaching consequences concerning nebular abundances throughout the universe.  However, NGC 7009 belongs to a minority group of planetary nebulae with large ADFs \citep[approximately 20\% of the total;][]{liu2010} that generally have densities and sizes implying that they are not very young \citep{robertsontessigarnett2005}.  The frequency of planetary nebulae with dust discs is similar \citep[17\%;][]{bilikovaetal2012}, though based upon a small sample.  On the other hand, most planetary nebulae and almost all \ion{H}{2} regions have ADFs of about a factor of 2 \citep[][]{garciarojasesteban2007, liu2010} and it is unclear whether the ADF phenomenon in these objects has the same origin as in the planetary nebulae with large ADFs \citep{garciarojasesteban2007}.   
Clearly, it would be worthwhile further investigating the kinematics and spatial distributions of ORLs in objects spanning a large range in ADF and nebular excitation.  

\section{Conclusions}\label{section_conclusions}

We analyze spatially- and velocity-resolved spectroscopy of NGC 7009 obtained with the UVES spectrograph at the ESO VLT (UT2) to study the kinematics of its emission lines.  We construct position-velocity maps for emission lines arising from different physical mechanisms and for the physical conditions.  

\begin{figure}
\begin{center}
\includegraphics[width=\columnwidth]{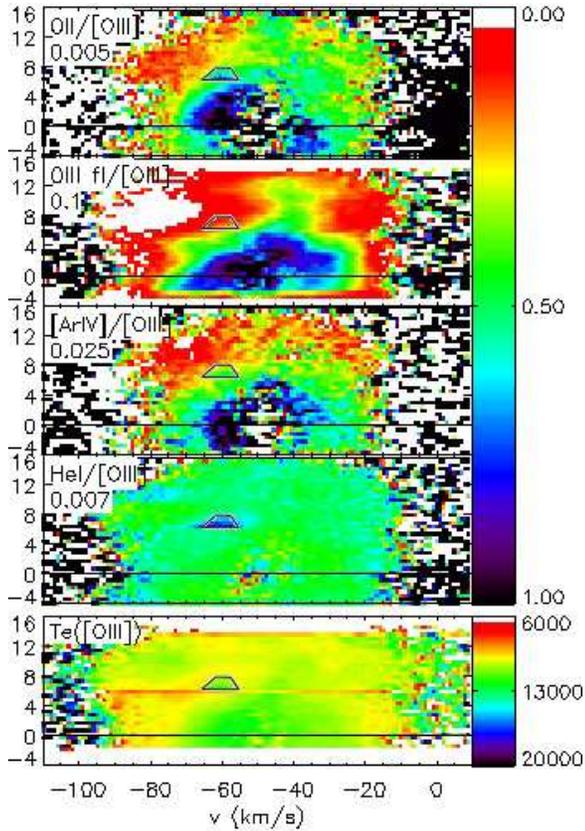}  
\end{center}
\caption{
These PV diagrams present ratios of \ion{O}{2} $\lambda\lambda$4639,4649 to [\ion{O}{3}] $\lambda$4959 (top), \ion{O}{3} $\lambda$3444 to [\ion{O}{3}] $\lambda$4959 (second), [\ion{Ar}{4}] $\lambda$4740 to [\ion{O}{3}] $\lambda$4959 (third), \ion{He}{1} $\lambda$4922 to [\ion{O}{3}] $\lambda$4959 (fourth), and the [\ion{O}{3}] temperature (bottom).  The scale factor by which each ratio was normalized is given in the top four panels.  Metal-rich clumps should appear as spatially-confined enhancements in the top panel.  Comparing the top two panels, the \ion{O}{2} $\lambda\lambda$4639,4649 emission arises from a much greater part of the O$^{2+}$ zone than does that from \ion{O}{3} $\lambda$3444.  
The first and fourth panels are very different, so the component emitting in \ion{O}{2} $\lambda\lambda$4639,4649 is not a strong emitter in \ion{He}{1} $\lambda$4922.  The trapezoids indicate the saturated region in the [\ion{O}{3}] $\lambda$4959 PV diagram.  The white level extends to 3\% of the scale factor (top four panels) and to 6,390\,K (bottom panel).  See Fig. \ref{fig_pv_diagrams} for further details.
}
\label{fig_pv_o2}
\end{figure}

We are able to study the ionization structure for oxygen in exquisite detail: O$^{3+}$ via \ion{O}{3} $\lambda$3707, the transition zone to O$^{2+}$ via \ion{O}{3} $\lambda$3444 (Bowen fluorescence) and \ion{O}{3} $\lambda\lambda$3757,5592 (charge transfer), the O$^{2+}$ zone via [\ion{O}{3}] $\lambda\lambda$4363,4959 (collisional excitation) and \ion{O}{2} $\lambda\lambda$4639,4649 (recombination), the O$^+$ zone via [\ion{O}{2}] $\lambda\lambda$3726,3729, and the outermost H transition zone via [\ion{O}{1}] $\lambda$6300.  The \emph{only} anomaly is that the kinematics of the collisionally-excited [\ion{O}{3}] $\lambda\lambda$4363,4959 lines does not coincide with that of the \ion{O}{2} $\lambda\lambda$4639,4649 recombination lines.  The kinematics favor at least two emission components for the \ion{O}{2} $\lambda\lambda$4639,4649 recombination lines.  The plasma from which a large fraction of the \ion{O}{2} emission arises clearly does not originate within the volume expected from the viewpoint of ionization equilibrium as traced by the [\ion{O}{3}] emission (\S\S \ref{section_ionization_structure} and \ref{section_location_recombination}, Fig. \ref{fig_pv_o2}).  Instead, it arises from a zone interior to it.  There is also a widely-distributed plasma component, co-spatial with the collisional [\ion{O}{3}] emission, that also gives rise to \ion{O}{2} $\lambda\lambda$4639,4649 recombination emission.  

The \ion{C}{2}, \ion{N}{2}, and \ion{Ne}{2} recombination lines appear to follow the distribution of the \ion{O}{2} lines.  (Their PV diagrams are usually of lower signal-to-noise.)  So, in NGC 7009, a large fraction of the \ion{C}{2}, \ion{N}{2}, \ion{O}{2}, and \ion{Ne}{2} recombination emission appears to arise internal to the zone expected to contain the doubly-ionized parent ions.  This result is in agreement with studies of the spatial distributions of recombination and forbidden lines (see \S \ref{section_location_recombination}) that typically find that the former are more centrally-concentrated.     

The kinematics of the \ion{C}{2}, \ion{N}{2}, \ion{O}{2}, and \ion{Ne}{2} recombination lines in NGC 7009 clearly favor a separate plasma component emitting mostly (only?) in these lines.  In NGC 7009, this plasma component would be metal-rich (as opposed to H-poor) since it is not a strong emitter in \ion{He}{1} lines.  We are unable to find unambiguous kinematic evidence for the existence of metal-rich clumps \citep[e.g.,][]{liuetal2000}, but cannot rule them out.  Other explanations for the origin of the \ion{C}{2}, \ion{N}{2}, \ion{O}{2}, and \ion{Ne}{2} lines, such as temperature fluctuations or modifications of atomic physics are either ruled out or implausible for NGC 7009 considering the kinematics we observe.  

The existence of multiple plasma components has very important consequences for the calculation and interpretation of chemical abundances.  Given the kinematics we find for NGC 7009, the discrepant abundances inferred from forbidden and recombination lines pertain to different plasma components within the nebula.  In NGC 7009, the \lq\lq abundance discrepancy" arises because these plasma components have different chemical compositions and are probed by emission lines excited by different mechanisms.  Until we understand the origin of these  plasma components, the interpretation of their chemical composition will be ambiguous.  There is no doubt that both components are present within the nebula in gaseous form.  However, both may not be relevant for all astrophysical purposes, particularly if these purposes pertain to the composition of the ejecta of the progenitor star.  

As regards the physical conditions and chemical abundances derived from collisionally-excited lines in NGC 7009, we find few problems that could prevent their use to reliably measure the chemical composition of the plasma component from which they arise.  The only surprise we find is that we directly observe a recombination contribution to the [\ion{N}{2}] $\lambda$5755 line that is often used as a temperature indicator in ionized nebulae.  Once this is accounted for, the [\ion{O}{3}], [\ion{Ne}{3}], and [\ion{N}{2}] temperatures agree.  The [\ion{Ar}{4}] and [\ion{O}{2}] densities present a coherent picture of the structure seen in direct images.  These results indicate that the atomic data for these species should be reliable.  We therefore conclude that forbidden lines provide reliable chemical abundances for the plasma component from which they arise.

Finally, we recommend further study of the kinematics of recombination lines to determine whether the kinematics we find for the \ion{C}{2}, \ion{N}{2}, \ion{O}{2}, and \ion{Ne}{2} lines in NGC 7009 is a common phenomenon.  Likewise, further study to determine the frequency of dust discs around the central stars of planetary nebulae would be very useful to ascertain whether these could be an explanation for the ADF phenomenon, at least in planetary nebulae with large ADFs \citep{bilikovaetal2012}.

\acknowledgments

We acknowledge financial support throughout this project from CONACyT grants 43121, 49447, 82066, 106719, and 129753 and from UNAM-DGAPA grants IN105511, IN108406, IN110011, and IN116908.  This research has made use of the SIMBAD database, operated at CDS, Strasbourg, France.  We thank Jos\'e Alberto L\'opez and the anonymous referee for helpful comments.


\begin{deluxetable}{lcc}
\tabletypesize{\small}
\tablecaption{Adopted line wavelengths\label{tab_wavelengths}}
\tablehead{
\colhead{Line} & \colhead{Wavelength} & \colhead{Splitting} \\
\colhead{} & \colhead{(\AA)} & \colhead{(km/s)} 
}
\tablewidth{3.15truein} 
\tablecolumns{3}
\startdata
optical recombination lines   &          & \\
H$\beta$                      & 4861.320 & 39.2 \\
H$\alpha$                     & 6562.791 & 39.4 \\
\ion{He}{1} $\lambda$4922     & 4921.931 & 40.1 \\
\ion{He}{2} $\lambda$4686     & 4685.748 & 33.1 \\
\ion{C}{2} $\lambda$4267      & 4267.261 & 35.4 \\
\ion{C}{2} $\lambda$6578      & 6578.05  & 36.0 \\
\ion{N}{2} $\lambda$5680      & 5679.56  & 35.1 \\
\ion{O}{2} $\lambda$4639      & 4638.856 & 36.6 \\
\ion{O}{2} $\lambda$4649      & 4649.135 & 35.5 \\
\ion{Ne}{2} $\lambda$3335     & 3334.837 & 35.1 \\
\ion{Ne}{2} $\lambda$3694     & 3694.215 & 38.0 \\
\ion{N}{3} $\lambda$4379      & 4379.11  & 34.1 \\
\ion{O}{3} $\lambda$3707      & 3707.24  & 34.0 \\
collisionally-excited lines   &          & \\
{[}\ion{N}{2}] $\lambda$5755  & 5754.57  & 37.8 \\
{[}\ion{N}{2}] $\lambda$5755 recomb.  & 5754.57  & 36.0 \\
{[}\ion{N}{2}] $\lambda$6583  & 6583.39  & 41.5 \\
{[}\ion{O}{1}] $\lambda$6300  & 6300.32  & \\
{[}\ion{O}{2}] $\lambda$3726  & 3726.05  & 40.4 \\
{[}\ion{O}{2}] $\lambda$3729  & 3728.80  & 41.7 \\
{[}\ion{O}{3}] $\lambda$4363  & 4363.21  & 37.1 \\
{[}\ion{O}{3}] $\lambda$4959  & 4958.92  & 39.2 \\
{[}\ion{Ne}{3}] $\lambda$3869 & 3868.76  & 38.2 \\
{[}\ion{Ar}{4}] $\lambda$4711 & 4711.34  & 36.0 \\
{[}\ion{Ar}{4}] $\lambda$4740 & 4740.20  & 35.7 \\
Bowen fluorescence lines      &          & \\
\ion{N}{3} $\lambda$4641      & 4640.64  & 36.2 \\
\ion{O}{3} $\lambda$3444      & 3444.10  & 34.8 \\
charge transfer lines         &          & \\
\ion{O}{3} $\lambda$5592      & 5592.37  & 35.2 \\
\ion{O}{3} $\lambda$3757      & 3757.21  & 34.9 \\
\enddata
\end{deluxetable}

\clearpage

\clearpage

\clearpage

\clearpage





\clearpage

\clearpage

\clearpage

\clearpage

\clearpage

\clearpage

\clearpage

\begin{thebibliography}{}

\bibitem[Aller (1984)]{aller1984} Aller, L. H. 1984, Physics of Gaseous Nebulae (Dordrecht, the Netherlands: D. Reidel Publishing Co.)

\bibitem[Barker (1982)]{barker1982} Barker, T. 1982, \apj, 253, 167

\bibitem[Barker (1991)]{barker1991} Barker, T. 1991, \apj, 371, 217

\bibitem[Barlow et al. (2006)]{barlowetal2006} Barlow, M. J., Hales, A. S., Storey, P. J., Liu, X.-W., Tsamis, Y. G., \& Aderin, M. E. 2006, in Planetary Nebulae in our Galaxy and Beyond, Proceedings IAU Symp. 234 (ed. M. J. Barlow \& R. H. M\'endez), 367

\bibitem[Bil\'\i kov\'a et al. (2012)]{bilikovaetal2012} Bil\'\i kov\'a, J., Chu, Y.-H., Gruendl, R. A., Su, K. Y. L., \& De Marco, O. 2012, \apjs, 200, 3

\bibitem[Bohigas (2009)]{bohigas2009} Bohigas, J. 2009, \rmxaa, 45, 107

\bibitem[Bohigas et al. (1994)]{bohigasetal1994} Bohigas, J., L\'opez, J. A., \& Aguilar, L. 1994, \aap, 291, 595

\bibitem[Bowen (1960)]{bowen1960} Bowen, I. S. 1960, \apj, 132, 1

\bibitem[Chu et al. (2011)]{chuetal2011} Chu, Y.-H., Su, K. Y. L., Bil\'\i kov\'a, J., Gruendl, R. A., De Marco, O., Guerrero, M. A., Updike, A. C., Volk, K., \& Rauch, T. 2011, \aj, 142, 75

\bibitem[Clegg et al. (1999)]{cleggetal1999} Clegg, R. E. S., Miller, S., Storey, P. J., \& Kisielius, R. 1999, A\&AS, 135, 359

\bibitem[Esteban et al. (2004)]{estebanetal2004} Esteban, C., Peimbert, M., Garc\'\i a-Rojas, J., Ruiz, M. T., Peimbert, A., \& Rodr\'\i guez, M. 2004, \mnras, 355, 229

\bibitem[Fang \& Liu (2011)]{fangliu2011} Fang, X., \& Liu, X.-W. 2011, \mnras, 415, 181

\bibitem[Fang \& Liu (2012)]{fangliu2012} Fang, X., \& Liu, X.-W. 2012, arXiv:1212.0005

\bibitem[Ferland et al. (2013)]{ferlandetal2013} Ferland, G. J., Porter, R. L., van Hoof, P. A. M., Williams, R. J. R., Abel, N. P., Lykins, M. L., Shaw, G., Henney, W. J., \& Stancil, P. C. 2013, \rmxaa, 49, 137

\bibitem[Fillipenko (1982)]{fillipenko1982} Fillipenko, A. 1982, \pasp, 94, 715

\bibitem[Garc\'\i a-Rojas \& Esteban (2007)]{garciarojasesteban2007} Garc\'\i a-Rojas, J., \& Esteban, C. 2007, \apj, 670, 457

\bibitem[Garnett \& Dinerstein (2001a)]{garnettdinerstein2001} Garnett, D. R., \& Dinerstein, H. L. 2001a, \apj, 558, 145

\bibitem[Gon\c{c}alves et al. (2003)]{goncalvesetal2003} Gon\c{c}alves, D. R., Corradi, R. L. M., Mampaso, A., \& Perinotto, M. 2003, \apj, 597, 975

\bibitem[Grandi (1976)]{grandi1976} Grandi, S. A. 1976, \apj, 206, 658

\bibitem[Henney \& Stasi\'nska (2010)]{henneystasinska2010} Henney, W. J., \& Stasi\'nska, G. 2010, \apj, 711, 881

\bibitem[Kaufman \& Sugar (1986)]{kaufmansugar1986} Kaufman, V., \& Sugar, J. 1986, J. Phys. Chem. Ref. Data 15, 321

\bibitem[Liu (2006)]{liu2006} Liu, X.-W. 2006, in Planetary Nebulae in our Galaxy and Beyond, Proceedings IAU Symp. 234 (ed. M. J. Barlow \& R. H. M\'endez), 219

\bibitem[Liu (2010)]{liu2010} Liu, X.-W. 2010, in New Vision 400: Engaging Big Questions in Astronomy and Cosmology Four Hundred Years after the Invention of the Telescope, eds. D. G. York, O. Gingerich, S.-N. Zhang, and C. L. Harper, Jr; also arXiv:1001.3715v2

\bibitem[Liu et al. (2006)]{liuetal2006} Liu, X.-W., Barlow, M. J., Zhang, Y., Bastin, R. J., \& Storey, P. J. 2006, \mnras, 368, 1959

\bibitem[Liu \& Danziger (1993)]{liudanziger1993} Liu, X.-W., \& Danziger, I. J. 1993, \mnras, 261, 465

\bibitem[Liu et al. (2001)]{liuetal2001} Liu, X.-W., Luo, S. G., Barlow, M. J., Danziger, I. J., \& Storey, P. J. 2001, \mnras, 327, 141

\bibitem[Liu et al. (1995)]{liuetal1995} Liu, X.-W., Storey, P. J., Barlow, M. J., \& Clegg, R. E. S. 1995, \mnras, 272, 369

\bibitem[Liu et al. (2000)]{liuetal2000} Liu, X.-W., Storey, P. J., Barlow, M. J., Danziger, I. J., Cohen, M., \& Bryce, M. 2000, \mnras, 312, 585

\bibitem[Liu et al. (2004)]{liuetal2004a} Liu, Y., Liu, X.-W., Luo, S.-G., \& Barlow, M. J. 2004, \mnras, 353, 1231

\bibitem[Luo \& Liu (2003)]{luoliu2003} Luo, S. G., \& Liu, X.-W. 2003 in IAU Symp. 209, Planetary Nebulae: Their Evolution and Role in the Universe, eds. S. Kwok, M. A. Dopita, \& R. Sutherland (Astronomical Society of the Pacific: San Francisco), 393

\bibitem[Luo et al. (2001)]{luoetal2001} Luo, S. G., Liu, X.-W., \& Barlow, M. J. 2001, \mnras, 326, 1049

\bibitem[McLaughlin \& Bell (1993)]{mclaughlinbell1993} McLaughlin, B. M., \& Bell, K. L. 1993, \apj, 408, 753

\bibitem[Mesa-Delgado et al. (2008)]{mesadelgadoetal2008} Mesa-Delgado, A., Esteban, C., Garc\'\i a-Rojas, J. 2008, \apj, 675, 389

\bibitem[Mesa-Delgado et al. (2009b)]{mesadelgadoetal2009b} Mesa-Delgado, A., Esteban, C., Garc\'\i a-Rojas, J., Luridiana, V., Bautista, M., Rodr\'\i guez, M., L\'opez-Mart\'\i n, L., Peimbert, M. 2009b, \mnras, 395, 855

\bibitem[Mesa-Delgado et al. (2009a)]{mesadelgadoetal2009a} Mesa-Delgado, A., L\'opez-Mart\'\i n, L., Esteban, C., Garc\'\i a-Rojas, J., \& Luridiana, V. 2009a, \mnras, 394, 693

\bibitem[Mesa-Delgado et al. (2012)]{mesadelgadoetal2012} Mesa-Delgado, A., N\'u\~nez-D\'\i az, M., Esteban, C., Garc\'\i a-Rojas, J., Flores-Fajardo, N., L\'opez-Mart\'\i n, L., Tsamis, Y. G., \& Henney, W. J. 2012, \mnras, 426, 614

\bibitem[Mesa-Delgado et al. (2011)]{mesadelgadoetal2011} Mesa-Delgado, A., N\'u\~nez-D\'\i az, M., Esteban, C., L\'opez-Mart\'\i n, L., \& Garc\'\i a-Rojas, J. 2011, \mnras, 417, 420

\bibitem[Nicholls et al. (2012)]{nichollsetal2012} Nicholls, D. C., Dopita, M. A., \& Sutherland, R. S. 2012, \apj, in press; also arXiv:1204.3880

\bibitem[Nussbaumer \& Rusca (1979)]{nussbaumerrusca1979} Nussbaumer, H., \& Rusca, C. 1979, \aap, 72, 129

\bibitem[Nussbaumer \& Storey (1981)]{nussbaumerstorey1981} Nussbaumer, H., \& Storey, P. J. 1981, \aap, 99, 177

\bibitem[Nussbaumer \& Storey (1984)]{nussbaumerstorey1984} Nussbaumer, H., \& Storey, P. J. 1984, A\&AS, 56, 293

\bibitem[Osterbrock (1989)]{osterbrock1989} Osterbrock, D. E. 1989, Astrophysics of Gaseous Nebulae and Active Galactic Nuclei (Mill Valley, USA: University Science Books)

\bibitem[Otsuka et al. (2009)]{otsukaetal2009} Otsuka, M., Hyung, S., Lee, S.-J., Izumiura, H., \& Tajitsu, A. 2009, \apj, 705, 509

\bibitem[Otsuka et al. (2010)]{otsukaetal2010} Otsuka, M., Tajitsu, A., Hyung, S., \& Izumiura, H. 2010, \apj, 723, 658

\bibitem[Peimbert (1967)]{peimbert1967} Peimbert, M. 1967, \apj, 150, 825

\bibitem[Peimbert \& Peimbert (2006)]{peimbertpeimbert2006} Peimbert, M., \& Peimbert, A. 2006, in Planetary Nebulae in our Galaxy and Beyond, Proceedings IAU Symp. 234 (ed. M. J. Barlow \& R. H. M\'endez), 227

\bibitem[Perinotto et al. (2004)]{perinottoetal2004} Perinotto, M., Sch\"onberner, D., Steffen, M., \& Calonaci, C. 2004, \aap, 414, 993

\bibitem[Phillips et al. (2010)]{phillipsetal2010} Phillips, J. P., Cuesta, L. C., \& Ramos-Larios, G. 2010, \mnras, 409, 881

\bibitem[Pradhan et al. (2011)]{pradhanetal2011} Pradhan, A. K., Nahar, S. N., Eissner, W. B., \& Montenegro, M. 2011, \baas, 43, 440

\bibitem[Ralchenko et al. (2011)]{ralchenkoetal2011} Ralchenko, Yu., Kramida, A., Reader, J. and NIST ASD Team (2011). NIST Atomic Spectra Database (version 4.1), [Online]. Available: http://physics.nist.gov/asd, National Institute of Standards and Technology, Gaithersburg, MD

\bibitem[Rauch (2003)]{rauch2003} Rauch, T. 2003, \aap, 403, 709

\bibitem[Robertson-Tessi \& Garnett (2005)]{robertsontessigarnett2005} Robertson-Tessi, M., \& Garnett, D. R. 2005, \apjs, 157, 371

\bibitem[Rodr\'\i guez \& Garc\'\i a-Rojas (2010)]{rodriguezgarciarojas2010} Rodr\'\i guez, M., \& Garc\'\i a-Rojas, J. 2010, \apj, 708, 1551

\bibitem[Rubin et al. (2002)]{rubinetal2002} Rubin, R. H., Bhatt, N. J., Dufour, R. J., Buckalew, B. A., Barlow, M. J., Liu, X.-W., Storey, P. J., Balick, B., Ferland, G. J., Harrington, J. P., \& Martin, P. G. 2002, \mnras, 334, 777

\bibitem[Sabbadin et al. (2004)]{sabbadinetal2004} Sabbadin, F., Turatto, M., Cappellaro, E., Benetti, S., \& Ragazzoni, R. 2004, \aap, 416, 955

\bibitem[Sharpee et al. (2004)]{sharpeeetal2004} Sharpee, B., Baldwin, J., \& William, R. 2004, \apj, 615, 323

\bibitem[Steffen et al. (2009)]{steffenetal2009} Steffen, W., Esp\'\i ndola, M., Mart\'\i nez, S., \& Koning, N. 2009, \rmxaa, 45, 143

\bibitem[Su et al. (2007)]{suetal2007} Su, K. Y. L., Chu, Y.-H, Rieke, G. H. et al. 2007, \apjl, 657, 41

\bibitem[Tsamis et al. (2003)]{tsamisetal2003} Tsamis, Y. G., Barlow, M. G., Liu, X.-W., Danziger, I. J., \& Storey, P. J. 2003, \mnras, 345, 186

\bibitem[Tsamis \& Walsh (2011)]{tsamiswalsh2011} Tsamis, Y. G., \& Walsh, J. R. 2011, \mnras, 417, 2072

\bibitem[Tsamis et al. (2008)]{tsamisetal2008} Tsamis, Y. G., Walsh, J. R., P\'equignot, D., Barlow, M. J., Danziger, I. J., \& Liu, X.-W. 2008, \mnras, 386, 22

\bibitem[Tsamis et al. (2011)]{tsamisetal2011} Tsamis, Y. G., Walsh, J. R., V\'\i lchez, J. M., \& P\'equignot, D. 2011, \mnras, 412, 1367

\bibitem[Villaver et al. (2002)]{villaveretal2002} Villaver, E., Manchado, A., \& Garc\'\i a-Segura, G. 2002, \apj, 581, 1204

\bibitem[Wesson et al. (2005)]{wessonetal2005} Wesson, R., Liu, X.-W., \& Barlow, M. G. 2005, \mnras, 362, 424

\bibitem[Wiese et al. (1996)]{wieseetal1996} Wiese, W. L., Fuhr, J. R., \& Deters, T. M. 1996, J. Phys. Chem. Ref. Data, Monograph 7

\bibitem[Wilson (1950)]{wilson1950} Wilson, O. C. 1950, \apj, 111, 279

\bibitem[Wyse (1942)]{wyse1942} Wyse, A. B. 1942, \apj, 95, 356

\bibitem[Zeippen et al. (1987)]{zeippenetal1987} Zeippen, C. J., Butler, K., \& Le Bourlot, J. 1987, \aap, 188, 251

\end{thebibliography}
\end{document}